\documentclass{aa}
\usepackage{natbib}
\graphicspath{{./fig/}}
\usepackage{graphicx}
\usepackage{txfonts}
\usepackage{caption}
\usepackage{subcaption}
\usepackage[monochrome]{color}

\usepackage[]{xcolor}
\definecolor{citecolor}{HTML}{195D95}
\definecolor{green}{HTML}{20C29D}
\definecolor{BrickRed}{HTML}{C41010}

\usepackage{amsmath,amssymb,latexsym}

\usepackage[T1]{fontenc}
\usepackage{titlesec}
\titleformat{\paragraph}[runin]
{\bfseries\scshape}{\theparagraph}{1em}{}

\begin{document}

\title{A Physical Background Model for the Fermi Gamma-ray Burst Monitor}


\author{
  B. Biltzinger
  \and
  F. Kunzweiler
  \and
  J. Greiner
  \and
  K. Toelge\thanks{Present address: Capgemini Nederland B.V., Reykjavikplein 1, 3543 KA Utrecht, Netherlands}
  \and
  J. Michael Burgess
}

\institute{Max-Planck-Institut für Extraterrestrische Physik, Giessenbachstrasse 1, 85748 Garching, Germany. \label{inst1}}

\date{Received 18 December, 2019; accepted 22 May, 2020}


\abstract{\textcolor{green}{We present the first physically motivated background model for the Gamma-Ray
   Burst Monitor (GBM) onboard the Fermi satellite. Such a physically
   motivated background model has the potential to significantly improve the 
   scientific output of Fermi/GBM, as it can be used to improve the background 
   estimate for spectral analysis and localization of Gamma-Ray Bursts (GRBs) and other sources.
   Additionally, it can also lead to detections of new transient events, 
   since long/weak or slowly rising ones do not activate one of the existing 
   trigger algorithms.
   In this paper we show the derivation of such a physically motivated 
   background model, which includes the modeling of the different background 
   sources and the correct handling of the response of GBM. While the goal
   of the paper is to introduce the model rather than developing a transient
   search algorithm, we demonstrate the ability
   of the model to fit the background seen by GBM by showing four applications,
   namely (1) for a canonical GRB, (2) for the ultra-long GRB 091024,
   (3) for the V404 Cygni outburst in June 2015, and (4) the ultra-long GRB 130925A.}}


\maketitle
%

\section{Introduction}
The background seen by astrophysical instruments can be very hard
  to model, as it normally contains a plethora
of unmodeled sources, instrumental background, and general
unknowns. Nevertheless, these background signals can contain
information which, when modeled, can be extracted for scientific
study. For instruments with the ability to spatially separate source
and background via e.g. focusing techniques, the characterization of
background signals can be simple, at least in the sense that spatially
distinct sources of background emission can readily be
identified. When an instrument lacks imaging capabilities in the
classical sense, i.e. its data are all-sky light curves with sources
superimposed upon each other, modeling and separation of background
sources must rely on temporal information such as the location of the
instrument or viewing direction with time.
Then, sources of background emission can be
modeled, and pushed through this temporal response of the
instrument. \\

An instrument which is part of the described class of non-imaging
instruments is the Gamma-ray Burst Monitor (GBM). GBM is one of two
instruments that are part of the Fermi Gamma-Ray Space Telescope which
was launched in 2008 into a low-Earth orbit. It consists of 12 sodium
iodide (NaI) and 2 bismuth germanate (BGO) detectors,
pointing in different directions (see Fig. \ref{fig:GBM}). GBM has
an all-sky coverage, except for the part that is occulted by the
Earth. The detectors nominally cover an energy interval from 8 keV to
1 MeV for the NaI detectors and 150
keV to 40 MeV for the BGO detectors \citep{2009ApJ...702..791M}.
GBM is used to detect and spectrally analyze transient events. It
provides an extension of the energy range over which transient events
can be observed below the the Large Area Telescope (LAT) energy
range (which covers 20 MeV to 300 GeV); also it allows fast on-board localization
of transients, to point the LAT towards
the transient event to detect potential high-energy delayed emission \citep{2009ApJ...702..791M}.
As a non-imaging instrument, GBM is inundated with an all-sky
background composed of several constant and temporally evolving
sources. Thus, for the identification of temporal transients such as GRBs,
GBM relies on the ability to separate source and background in any analysis.
The classical approach in estimating the background temporally and
spectrally is to identify off-source regions in time and fit smooth
functions such as polynomials to each Pulse Height Analysis (PHA)
channel's temporally evolving signal. Then, the estimated model is extrapolated 
into the on-source region, thus providing an estimate for
the background in that time interval \citep{1999ApJ...512..362P, Greiner_2016}.\\

This classical approach is suitable for transient events which are
short with respect to the typical temporal variation of the background
(about 10 min.), but can be very problematic for long duration events, as the
background extrapolation can be poor and inaccurate due to
non-polynomial variations in the signal. Moreover, assuming
independent temporal variations for each channel allows for a high
degree of freedom that is unlikely to exist. An alternative approach is
to construct a background model from physical components, that evolve temporally and spatially.
Such a model can be complex, due to the many different
background components, which explains that after 12 years of operation of GBM and 28 years after the
CGRO/BATSE observations, there is still no properly working physical model to describe the
background of GBM.

Some empirical attempts have been made in the past:
(1) \citet{2013A&A...557A...8S} fit the
background around trigger times of GBM with separate polynomials (up to third
order) to different quantities like the angle between the detector and
the Earth, forming a geometrical background model.
But this approach gives the
fit a lot of freedom as it has many free unbound parameters, introduces
ambiguity between the different polynomials and does
not use physical information of the sources that create the background.
Another work that used the background seen by GBM was done by (2) \citet{2015arXiv150203399H}.
In this work they searched for decay lines of sterile neutrinos in the background
spectrum of GBM. They used data cuts to get rid of most of the background
contamination from e.g. cosmic rays and the Earth Albedo, but
did not try to model all the background sources.
(3) \citet{Fitzpatrickbackground} used the fact that the detectors of GBM, in most cases, are at the same
    geographical coordinates and point to the same position in the sky every 30th orbit. Thus, they assume that the background at time T should be the same as at time T $\pm$ 30 orbits. This is a purely empirical method that for example breaks down, like stated in \citet{Fitzpatrickbackground},
    when there is an Autonomous Repoint Request (ARR) or when due to varying solar activity the particle
    flux changes or the Earth's magnetic field is compressed.
    An ARR causes the Fermi satellite to slew, to point the LAT towards a GRB. If this happens, the pointing direction of the GBM detectors will not be the same 30 orbits later.

We wanted to build a predictive background model with a minimum number of free parameters,
that incorporates the physics of the source types and the response of the GBM detectors.
Such a model
should only fit the background and not also (long or slowly varying) transient signals,
and should allow us to determine
the individual contributions of the different background sources.
The model presented herein treats all its components physically except the cosmic ray background, which is explained in subsection \ref{cosmic_rays}.
One possible future application for a physically motivated background model could be the possibility
to detect events with slow raising count rates (e.g. ultra-long
GRBs). Herein, we show some examples of the background seen by GBM in Sec. \ref{gbm_and_background},
describe the setup up of such a physically motivated model in Sec. \ref{sec:model},
and show results for the ultra-long GRB 091024, the 2015 V404 Cygni outburst and GRB 110920A in Sec. \ref{sec:results}.

\section{GBM and its Background}
\label{gbm_and_background}

GBM converts detected photon events into 128 Pulse Height Analysis
(PHA) channels according to the 'height' of the electronic signal at
the end of the photo-multiplier tube.  These signals are then stored as
single events with time tags and the corresponding PHA channel number
in the Time-Tagged Events (TTE) data files, but also binned in 8 second time bins with full
spectral resolution of 128 PHA channels in the CSPEC data files and
with 256 ms time bins with
a reduced spectral resolution of only 8 PHA channels in the CTIME data files.
The continuous TTE data exist only since 2010 \citep{2009ApJ...702..791M}.
\\
\begin{figure}[h]
  \centering
  \resizebox{\hsize}{!}{\includegraphics{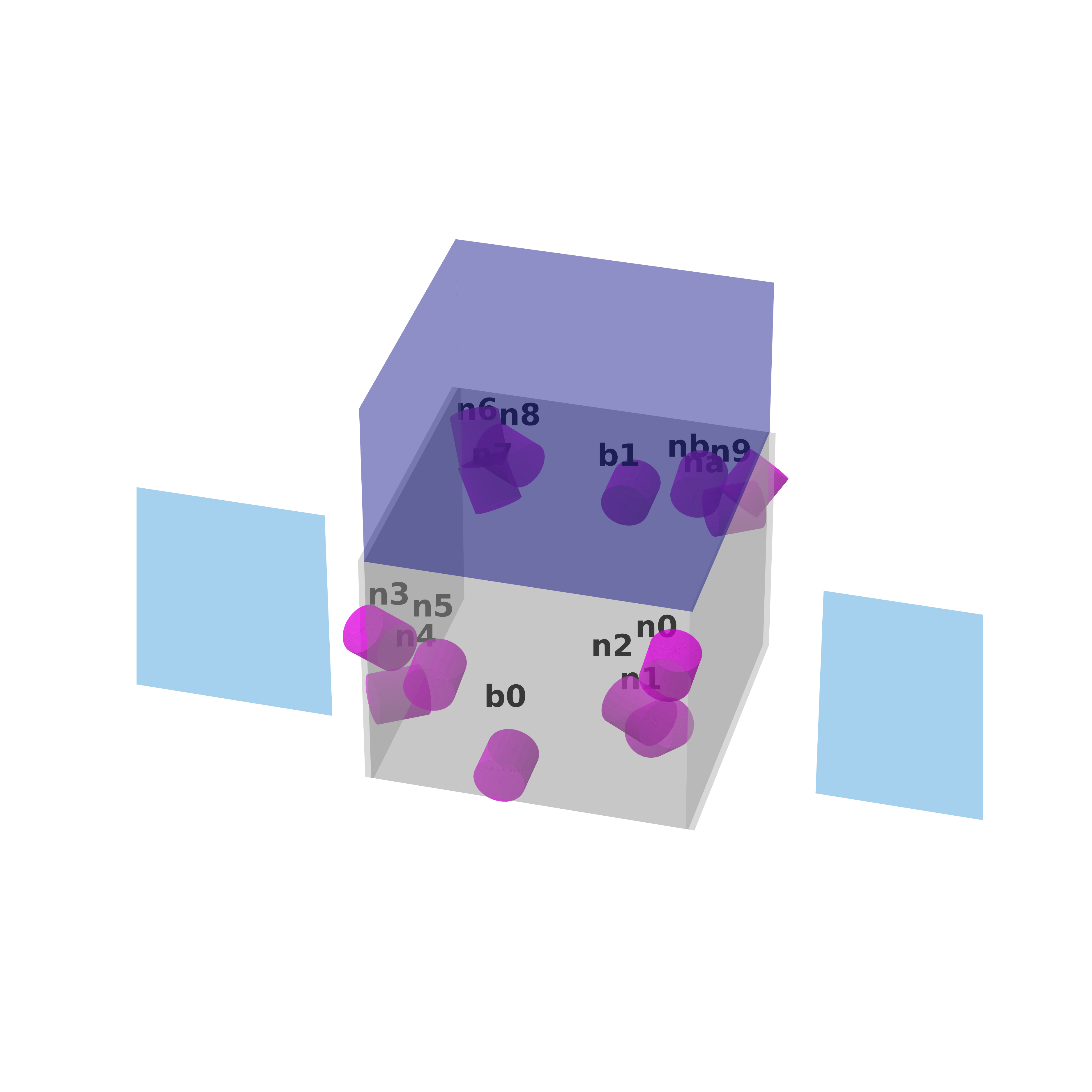}}
  \caption{Schematic picture of the setup of GBM on board of the Fermi satellite. It shows the mounting points and pointing directions for all 14 detectors that are part of GBM. Additionally the satellite base is colored gray, the LAT dark blue and the solar panels light blue.}
  \label{fig:GBM}
\end{figure}

\begin{figure}[h]
  \centering
  \includegraphics*[]{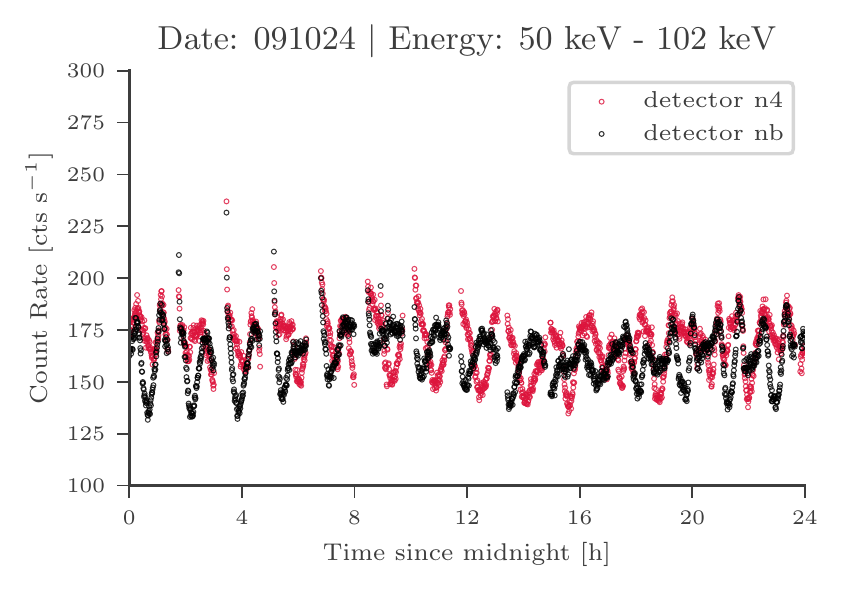}
  \caption{Background variation at a binning of 15 seconds for the date 24/10/2009 for the
    two detectors n4 (red circles) and nb (black circles)
    covering the same reconstructed energy range 50 keV to 102 keV. The time
    variation of the background is completely different for the
    two detectors.}
  \label{fig:echan_4_nb_and_n4}
\end{figure}

\begin{figure}[h]
  \centering
  \includegraphics*[]{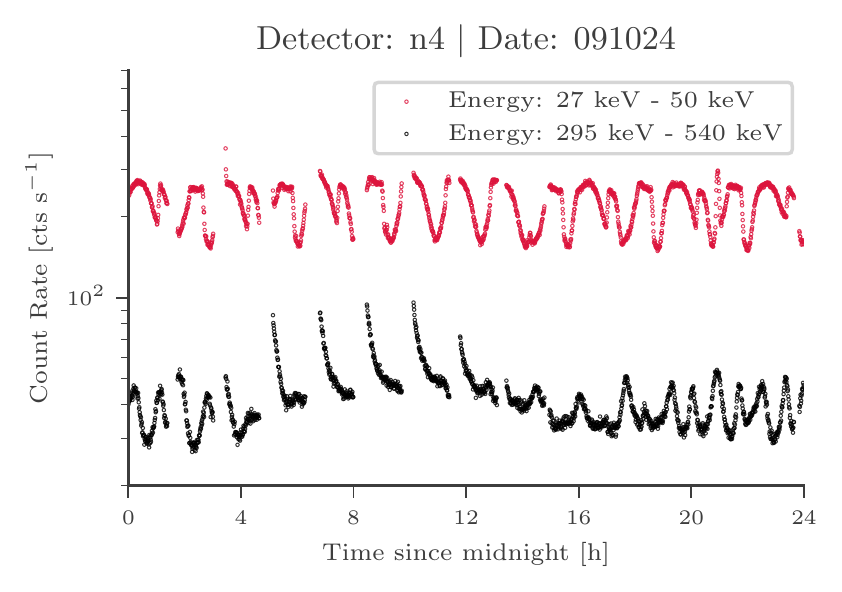}
  \caption{Background variation at a binning of 15 seconds for the date 24/10/2009 for the
    detector n4 and for the two reconstructed energy ranges 27 keV to 50 keV
    (red circles) and 295 keV to 540 keV (black circles). The
    time variation of the background is completely different for
    the two energy ranges.}
  \label{fig:det_n4_echan_2_and_5-min.pdf}
\end{figure}

The background seen by the individual detectors of GBM (see Fig. \ref{fig:GBM}) is a superposition
of several different background sources, whose contributions to the
total background are strongly dependent on the energy and the
orientation of the detectors with respect to the position of the
background sources (more details about the different background sources
will be described in Sec. \ref{bkg_sources}). Since the off-axis sensitivity
is energy-dependent, the same photon spectrum will result in different count
spectra depending on the orientation of the detector with respect to the
source position. The dependence on the
orientation leads to different background variations for different
detectors in the same reconstructed energy range, while the dependence on the
energy leads to different background variations for the same detector but
different reconstructed energy ranges. This is shown in
Fig. \ref{fig:echan_4_nb_and_n4} and \ref{fig:det_n4_echan_2_and_5-min.pdf}.
The background model needs to be capable to explain these different
backgrounds for different detector orientations and PHA channels.

\section{Model}
\label{sec:model}
\subsection{Response}
\begin{figure*}
  \centering

  \begin{subfigure}[b]{0.45\textwidth}
    \centering
    \includegraphics[width=\textwidth]{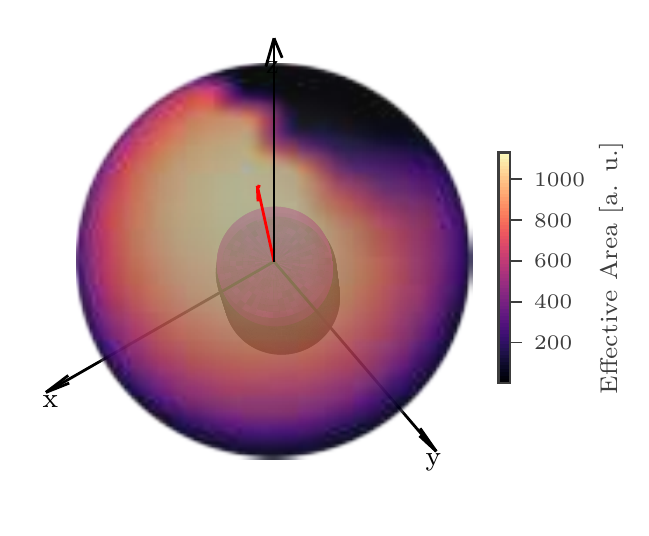}
    \label{fig:shielding_a}
  \end{subfigure}
  \hfill
  \begin{subfigure}[b]{0.45\textwidth}
    \centering
    \includegraphics[width=\textwidth]{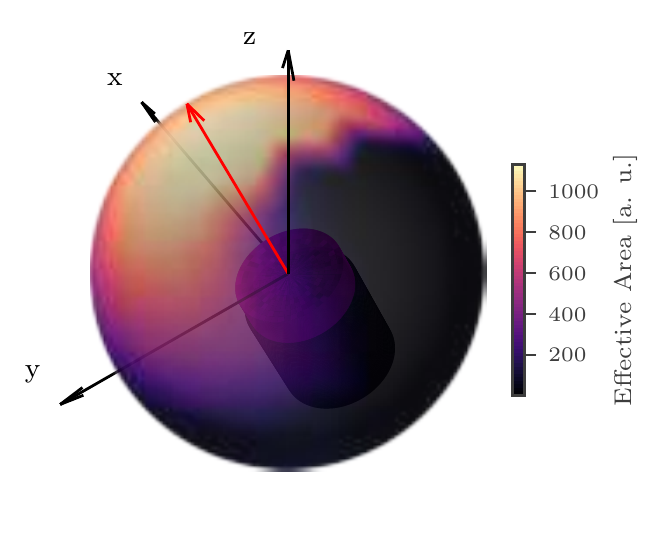}
    \label{fig:shielding_b}
  \end{subfigure}
  \begin{subfigure}[b]{0.45\textwidth}
    \centering
    \includegraphics[width=\textwidth]{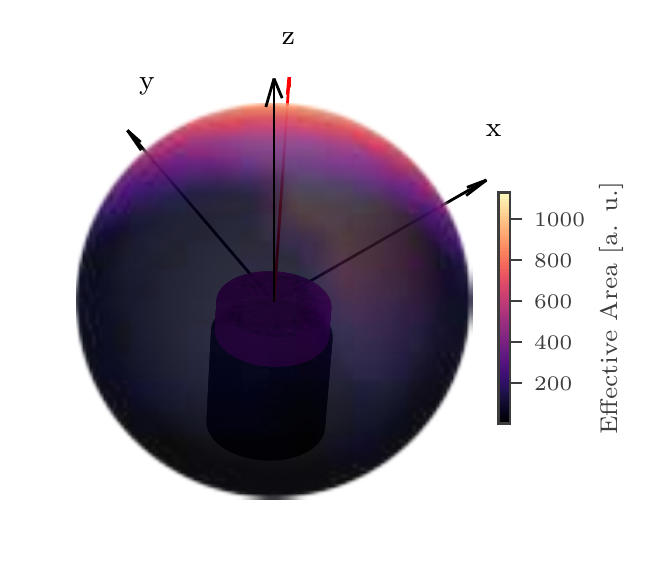}
    \label{fig:shielding_c}
  \end{subfigure}
  \hfill
  \begin{subfigure}[b]{0.45\textwidth}
    \centering
    \includegraphics[width=\textwidth]{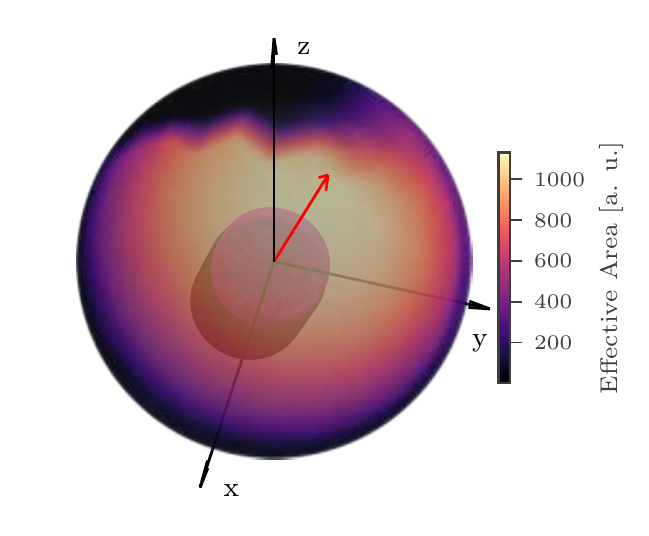}
    \label{fig:shielding_d}
  \end{subfigure}

  \caption{Effective area seen by a photon with energy 38 keV to be
    detected in the reconstructed energy channel 2 (CTIME) for different arrival
    directions on a unit sphere around detector n0 for four
    different viewing directions on the sphere. The red arrow
    marks the line of sight of the detector n0 (optical axis),
    the dark cylinder symbolizes the detector housing and the
    magenta cylinder the detector crystal. The influence of the
    LAT is clearly visible in the upper left plot, as it cuts
    away the effective area on one side of the line of
    sight. The coordinate system is defined such that the z-axis is the pointing direction of the LAT and the x-axis points along the normal of the satellite side on which the b0 detector is mounted. Responses generated with \texttt{gbm\char`_drm\char`_gen} \citep{2018MNRAS.476.1427B, burgess_james_michael_2019_2590555, 2019ApJ...873...60B}.}
  \label{fig:shielding}
\end{figure*}

For spectral instruments with large energy dispersion like GBM, modeling the
response is crucial for any physical interpretation of the data. The
response gives the effective area seen by a photon with a given
energy $E_{ \mathrm{ph}}$ to be detected in one of the detector's PHA
channels $E_{ \mathrm{det}}$, and therefore connects the physical
spectrum of the source with the count rates in the PHA channels.  The
effects that are incorporated in the response include on the single
detector level the geometry of the scintillation crystal, energy
dispersion, partial energy deposition of the photons in the crystal
and absorption of photons by the detector housing and photo-multiplier
tubes attached to the crystal. Additionally, on the satellite level,
shielding by other components of the satellite and other detectors is
taken into account (see Fig. \ref{fig:shielding}). The response is
therefore a function of the position of a source in the satellite
frame, the energy of
the incoming photon and the PHA channel in which the photon is detected \citep{article}.\\

Each PHA channel has an associated reconstructed energy range, but due
to energy dispersion and partial energy deposition of the photon in
the crystal this does not imply that all photons detected in this PHA
channel have a physical energy within the
reconstructed energy range associated to this PHA channel.\\

The connection of the physical spectrum $F$ [photons cm$^{-2}$ s$^{-1}$]
and the detected count rate spectrum $D$ [counts s$^{-1}$] via the
response $R$ is given in Eq. \ref{response_general}.
\begin{equation}
  D(E_{ \mathrm{det}}) = \int  \,  \mathrm{d} E_{ \mathrm{ph}} \, R(E_{ \mathrm{ph}},  E_{ \mathrm{det}}, \phi_{\mathrm{source}}, \theta_{ \mathrm{source}}) \cdot \, F( E_{ \mathrm{ph}})
  \label{response_general}
\end{equation}

\noindent The quantities $\phi_{ \mathrm{source}}$ and $\theta_{ \mathrm{source}}$
define the position of the source in the satellite frame, with $\theta_{
  \mathrm{source}}$ being the zenith angle and +$\phi_{ \mathrm{source}}$
the azimuth angle measured from the pointing direction of the b0 detector.

\begin{figure}
  \centering
  \includegraphics*[]{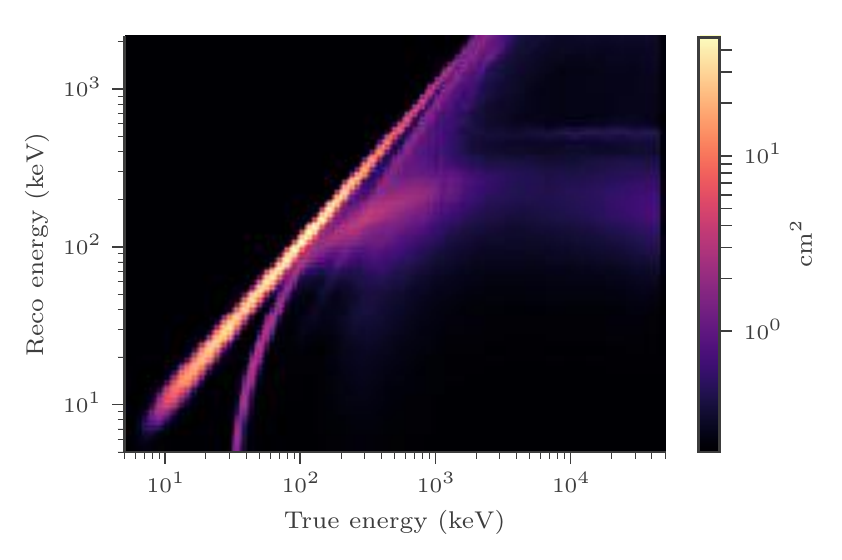}
  \caption{Example response that shows the effective area for
    different photon energies to be detected in the PHA channels with
    associated reconstructed energy. Figure created with the
    \texttt{gbm\char`_drm\char`_gen} package \citep{2005NCimC..28..797H, burgess_james_michael_2019_2590555}.}
  \label{fig:DRM}
\end{figure}

To determine the detector response matrix (DRM), an on-ground
calibration of GBM was performed \citep{2009ExA....24...47B}. We used
the Python package \texttt{gbm\char`_drm\char`_gen}\footnote{\url{https://github.com/mpe-heg/gbm_drm_gen/}} to get the DRMs. The package
\texttt{gbm\char`_drm\char`_gen} uses the responses generated via simulation
to match the calibration of GBM. With its help one can calculate the DRMs for user
defined energy bins of the incoming photons, detector PHA channels
and positions of the source (see Fig. \ref{fig:DRM}) \citep{2018MNRAS.476.1427B, burgess_james_michael_2019_2590555, 2019ApJ...873...60B}.\\

\subsection{Response Implementation}
\label{subsec:response_implementation}

In the physical background model the temporally changing responses for
point sources (e.g the Sun) and for extended sources (e.g. Cosmic
Gamma-ray Background) are calculated to get the correct influence of
these sources on the count rates. The implementation of the response
for a point sources is, with the help of the \texttt{gbm\char`_drm\char`_gen} package,
straightforward:
\begin{enumerate}
\item Calculate the position of the point source in the satellite
  frame for several times with an appropriate time resolution
  (computation time vs. accuracy of interpolation in step 4).
\item Use the \texttt{gbm\char`_drm\char`_gen} package to get the DRMs for the different
  positions.
\item Convolve the assumed spectrum with the DRMs to get the expected
  count rates in the different PHA channels.
\item Interpolate between the times for which the point source
  position was calculated.
\end{enumerate}
The use of an interpolation in time in the last step will of course introduce a systematic error in our analysis. In our implementation, 800 time steps are used for one day which results in one time step every 108 seconds. In this short time the direction of the point source will move only very little in the satellite coordinate frame. Therefore the response change between two time steps is very small which is why the introduced systematic error is small. There are some regions of the GBM response with fast changes for similar directions of the incident photon, which are caused by some parts of the satellite like the LAT radiators. This may introduce a larger error at these times, but we checked that the results do not change if we make the time steps smaller.
For the extended sources this procedure has to be modified, as the
response is a function of the position of the source in the satellite
frame and extended sources cover a range of positions. The formula to
connect the flux $F$ [photons cm$^{-2}$ s$^{-1}$ sr$^{-1}$] of the
extend sources and the detected count rate spectrum $D$ [counts
s$^{-1}$] is given by
\begin{equation}
  \begin{split}
    D(E_{ \mathrm{det}}) = \int \int \, &  \mathrm{d}E_{ \mathrm{ph}} \,  \mathrm{d}\Omega \, R(E_{ \mathrm{ph}}, E_{ \mathrm{det}}, \phi, \theta) \cdot \, F(E_{ \mathrm{ph}}) \cdot \, f(\phi, \theta),
  \end{split}
  \label{response_extended}
\end{equation}
\noindent
where $f(\phi, \theta)$ is the occlusion function
\begin{equation}
  f(\phi, \theta) =
  \begin{cases}
    1, & \text{if the point is occulted by the source} \\
    0, & \text{otherwise.}
  \end{cases}
  \label{occlusion}
\end{equation}
For a point source $f(\phi, \theta)=\delta(\phi - \phi_{\mathrm{source}})\delta(\theta-\theta_{\mathrm{source}})$, with
which Eq. \ref{response_extended} simplifies to Eq.\ref{response_general}.\\

To get the expected count rate spectrum of
an extended source, the following procedure is applied:
\begin{enumerate}
\item Build a grid with $N_{ \mathrm{grid}}$ equally distributed points on a unit sphere
  around the detector in the satellite frame.
\item For each of these points, calculate the DRM with the \texttt{gbm\char`_drm\char`_gen} package.
\item Calculate the region covered by the source in the
  satellite frame for several times with appropriate time
  resolution (balancing computation time with accuracy of the
  interpolation in step 6).
\item Sum the responses of the relevant grid points and
  multiply the result by the solid angle that every point
  covers
  $\left(\frac{4\pi}{N_{ \mathrm{grid}}} \mathrm{sr} \right)$
  to get an effective response.
\item Convolve the assumed spectrum through the effective
  response to get the expected count rates in the different
  PHA channels.
\item Interpolate between the times for which the count rate was calculated.
\end{enumerate}
With the response effects for point sources and extended sources implemented,
we can predict the count rate spectrum for any given physical spectrum of any source. \\
    The intense flux of photons from a GRB scatters of the Earth's atmosphere and produce a secondary, weaker flux of photons which can be detected by GBM.
    Indeed all high-energy radiation can produce such a secondary flux, but at levels far below that of a GRB; therefore, we neglect atmospheric scattering
    from non-GRB sources in this work.

\subsection{Background sources}
\label{bkg_sources}
Our model incorporates 6 components which will be describes in this subsection.
We start with the simplest
model of a constant background and then adding more components one at a
time. For each step we show a fit of the model with the introduced background
components to the data of detector n6 for 26/10/2009 between 102 and 295 keV
reconstructed energy, showing
the improvement we can get by adding the different components to the model.
The used components can be categorized into two main classes:
The photon and the charged particle background components.  The first class is
characterized by the components producing a photon spectrum, which is
measured by the GBM detectors directly. Members of this class are the
Earth Albedo, the Cosmic Gamma-ray Background (CGB), point sources as
well as the Sun.  In the second class, containing cosmic rays and the
South Atlantic Anomaly, charged particles alter the background due to
excitation of the satellites atoms or via direct energy deposit in the
detectors.

\subsubsection{Constant Background} \label{parag:model_const}
\begin{figure}[h]
  \centering
  \includegraphics*[]{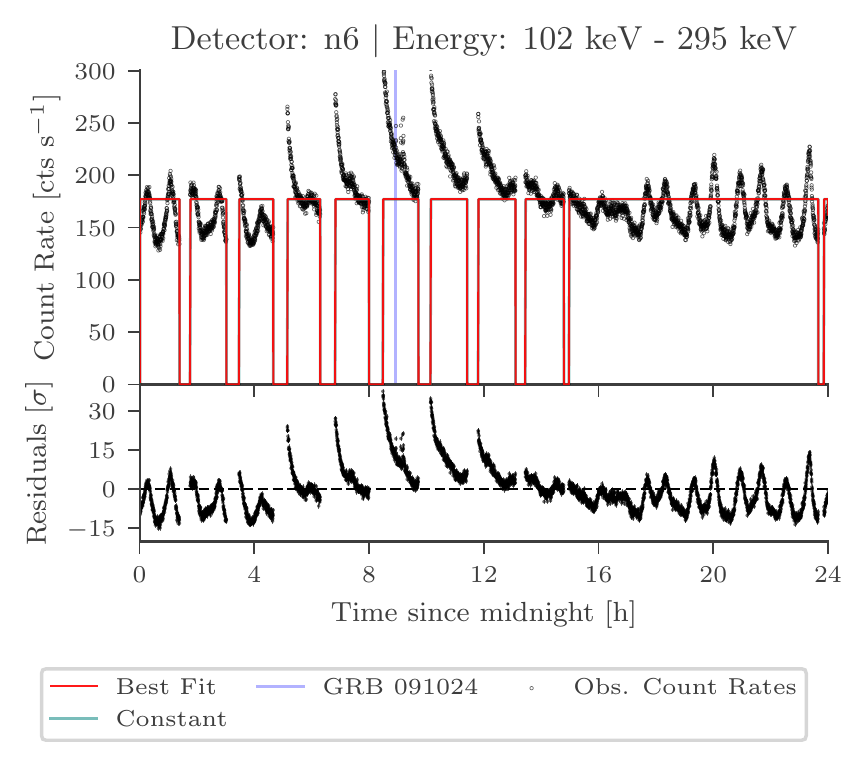}
  \caption{Fit of a constant line to the data of detector n4 and
    the energy of 295 keV - 540 keV of day 26/01/2015. The count
    rate of the model has been set to zero during the SAA
    passage where the detectors are turned off.
  The GBM trigger time for GRB 091024 is marked by a cyan line.}
  \label{fig:model_const}
\end{figure}

The most intuitive model is that of a constant background which is shown in
Fig. \ref{fig:model_const} plus gaps during every South Atlantic Anomaly (SAA)
transition when the detectors are shut down.\\

When building a physically motivated background model for GBM, for example in order to detect new sources,
it is reasonable to start by fitting for the sharply peaked and rapidly decaying
signal that occurs after the exits of the South Atlantic Anomaly. This
is our first extension to the constant background model.

\subsubsection{South Atlantic Anomaly}

There are two regions in the Earth's magnetic field which have the
right properties to permanently trap large densities of charged
particles (mostly protons and electrons) in radiation belts. These two
regions are called inner and outer Van-Allen Belts and both lie, outside of the SAA,
well above Fermi's orbit, which has an altitude of about 550 km, with the
inner Van Allen belt being located between 1000 and
6000km \citep{1959Natur.183..430V, da_Silva_2017,
  2011JGRA..116.9234G}.\\

The South Atlantic Anomaly (SAA) is a region located over the South
Atlantic, where the Earth's magnetic field is significantly weaker
\citep{doi:10.1002/2016SW001371}. This originates from an offset of
the dipole which approximates Earth's magnetic field and causes the
inner Van Allen belt to bend towards the Earth at the position of the
SAA \citep{2007PhDT.......265M} and reaching an
altitude as low as $100\,km$ that therefore intersects Fermi's orbit at $550\,km$.
In order to protect the detectors from damage caused by the high flux of charged
particles, the detectors are routinely shut down when the satellite crosses the
SAA.  Nevertheless, the satellite and detector material undergo
nuclear excitation by collision with the charged particles and subsequently photons are
produced by the de-excitation of the activated material which are
measured when the detectors are turned on again. As the high count
rates after the SAA originate from nuclear de-excitation, its influence
should decay exponentially over time and should have the functional
form
\begin{equation}
  R_{\mathrm{SAA}}(t) = R_{\mathrm{SAA}}(t_0) \cdot  \exp(-C_{\mathrm{decay}} \cdot (t-t_0)).
\end{equation}

\begin{figure}[h]
  \centering
  \includegraphics*[]{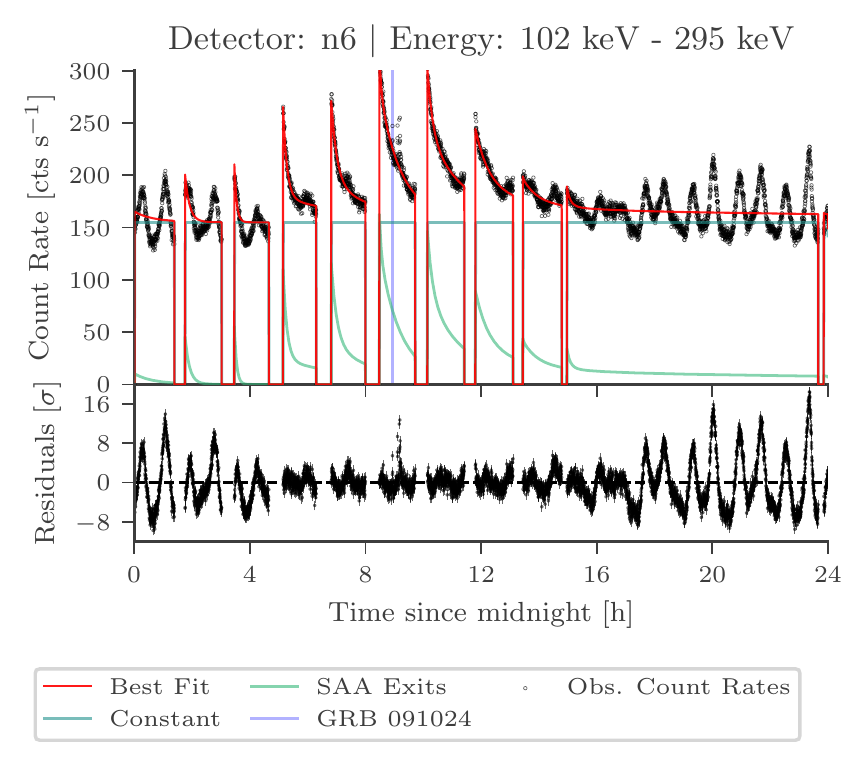}
  \caption{Same as Fig. \ref{fig:model_const}, but including the
    additional SAA component. The peaks in the residuals after the SAA exit have disappeared.
    The GBM trigger time for GRB 091024 is marked by a cyan line.
}
  \label{fig:model_const_saa}
\end{figure}

Because of the different elements in the satellite's and the detector's
material, there should be a superposition of several exponential
decays after each SAA exit.
In order to keep the computational effort
manageable, we included only two decay functions after each exit to allow for
a fast and a slow decay. The fast decaying part usually decays in $\approx$ minutes,
and is thus completely gone before the next SAA passage.
But the slowly decaying part can take several hours to decay completely which causes accumulation
of background signal by consecutive SAA passages. This activation decays slowly during the long time
of the day with no SAA passage. This leads to four parameters per SAA exit, one for each
decay constant and the two corresponding normalizations.
It is clearly visible in Fig. \ref{fig:model_const_saa} that the
model is able to explain the rapidly decaying count rates after the
SAA passage, and is significantly reducing the peaks in the residuals.

The presented procedure for the SAA influence depends on the boundaries, which are used by the Fermi team, being correct. If the boundaries of the SAA are wrong it can happen, that the detectors are for example turned off a bit too late, which leads to a rise in the count rates before the official SAA entrance. \\
As the SAA shape and dimension changes due to solar activity, and is also slowly moving over the years, the Fermi team is adapting the SAA boundaries as needed, so that such outliers should be very small in number. To be on the safe side one could always ignore some time (few hundreds of seconds) before and after the official SAA transition.
\subsubsection{Earth Albedo}
\label{earth_albedo}
\begin{figure}[h]
  \centering
  \includegraphics*[]{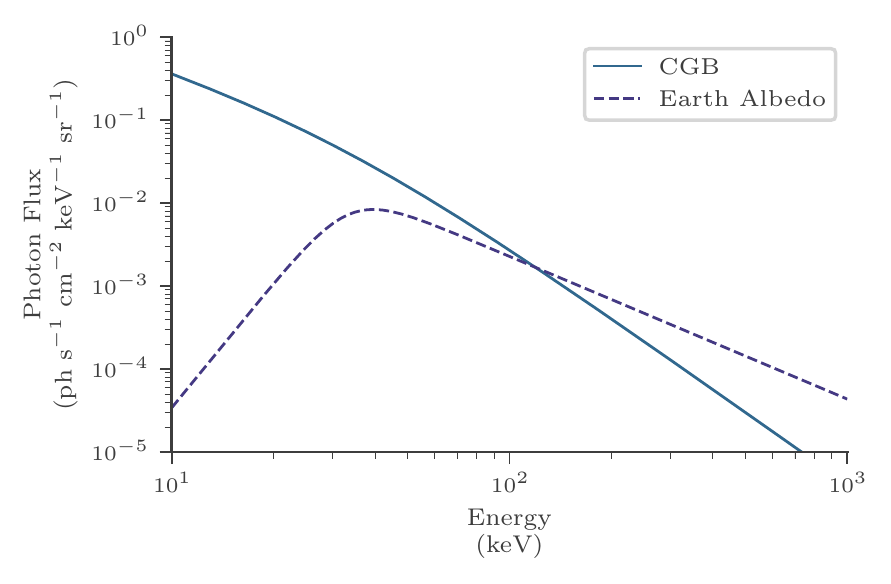}
  \caption{The spectra of the Cosmic Gamma-Ray Background (CGB) and
    the Earth Albedo for spectral parameters given in
    \citep{Ajello:2008xb}. The spectrum of the CGB is higher than the
    one of the Earth Albedo for energies smaller than $\approx$ 120
    keV, but for higher energies the spectrum of the CGB has a steeper
    slope which causes that the Earth Albedo to surpass the CGB
    spectrum for energies higher than $\approx$ 120 keV.}
  \label{fig:earth_albedo_spec}
\end{figure}

The Earth is known to be a source of $\gamma$-rays in the GBM energy range
\citep{1981JGR....86.1265T, Ajello:2008xb}. These $\gamma$-rays are produced
by cosmic rays colliding with molecules in the Earth's atmosphere
through $\pi^0$-decay and bremsstrahlung \citep{Petry:2004bb, 2009PhRvD..80l2004A}. The spectrum of the produced $\gamma$-rays that are
emitted back into space is called 'Earth Albedo' 
and is shown in Fig. \ref{fig:earth_albedo_spec} in comparison to the
Cosmic Gamma-Ray spectrum.

The secondary particles produced in the collision of a cosmic ray with
an atmospheric molecule tend to fly in the same direction as the
primary cosmic ray because of momentum conservation. As cosmic rays
mostly consist of protons, they get deflected eastwards upon impinging
Earth's magnetic field, therefore the cosmic ray flux from the west as
seen by the satellite is larger. It has been shown by \citet{Petry:2004bb} that the East-West
asymmetry in the Earth Albedo peaks at a photon energy in the GeV
range and is small at lower energies in the MeV
range. \citet{2009PhRvD..80l2004A} showed that the effect is
also very small for very high energies ($\approx$ TeV).
Additionally, nearly the whole photon flux with very high energies
($\approx$ TeV) originates from a small ring around the Earth (as seen by
the satellite).  This 'ring effect' weakens towards lower $\gamma$-ray energies,
therefore we assume that in the keV to low MeV region this effect can
be neglected \citep{2009PhRvD..80l2004A}.\\
It is therefore reasonable
to assume that the East-West and the ring
effect, that are well observed at higher energies, is not significant
in the keV to low MeV energy region and therefore to assume that the
Earth Albedo at these energies is isotropic over the whole Earth
surface seen by the satellite.

\begin{figure}[h]
  \centering
  \includegraphics*[]{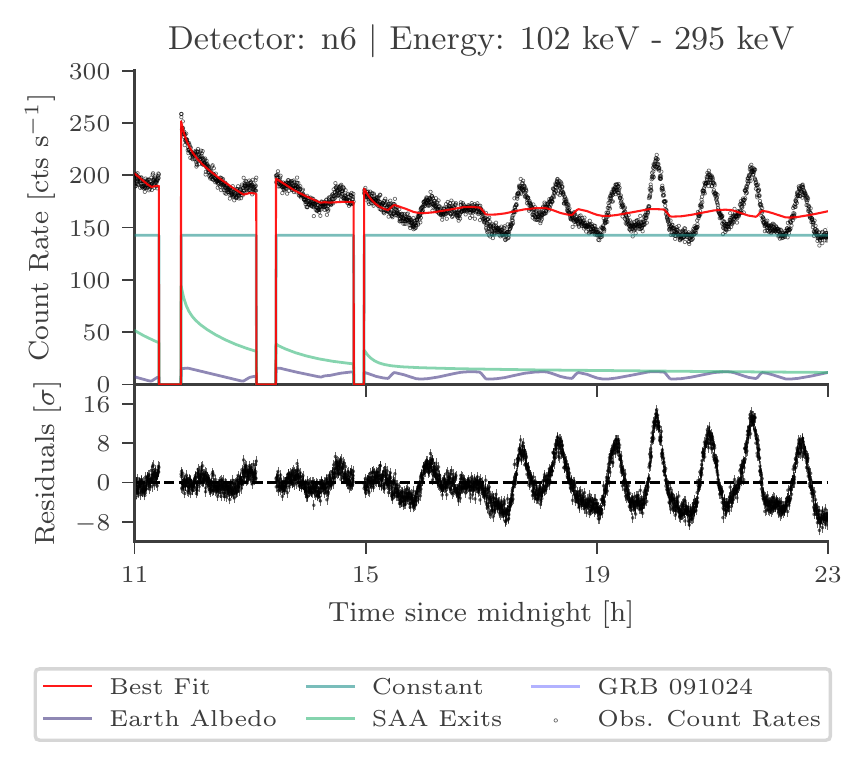}
  \caption{Same as Fig. \ref{fig:model_const_saa} but only for half of the day for better legibility and including the Earth Albedo as additional background source.}
  \label{fig:counts_earth_albedo}
\end{figure}

The variation of the background from the Earth Albedo in the
individual detectors is caused by the different positions of the Earth
in the satellite coordinate system for different times and the therefore
changing effective response (see Sec. \ref{subsec:response_implementation}). Fig.
\ref{fig:counts_earth_albedo} shows the model expanded by the
component for the Earth Albedo. The assumed functional form of the Earth Albedo
spectrum in this paper is a smoothly connected double power law with the spectral
parameter values as given in \citet{Ajello:2008xb},
but we free the normalization $C$ and fit it to the GBM data (see Eq. \ref{eq:earth_spec}).

\begin{equation}
  \label{eq:earth_spec}
  \frac{\mathrm{d}N}{\mathrm{d}E} = \frac{C}{ ( \frac{E}{33.7 \mathrm{keV}} ) ^{-5} + ( \frac{E}{33.7 \mathrm{keV}} ) ^{1.72}}
\end{equation}

\subsubsection{Cosmic Gamma-Ray Background}


The Cosmic Gamma-Ray Background (CGB), which was discovered more than 50
years ago, is generally agreed to be
produced by emission of unresolved extra-galactic point sources
\citep{PhysRevLett.9.439, 2004NuPhS.132...86H, Ajello:2008xb}. The galactic contribution to the CGB is subdominant for photon energies below 1 MeV. \citep{SPI_diffuse_galactic,Ajello:2008xb} and almost the entire Cosmic X-ray background (CXB) for energies smaller than 2 keV is accounted for by Active Galactic Nuclei (AGN) that host accreting super-massive black holes
\citep{2004NuPhS.132...86H}.  For energies above 6 keV on the other hand,
the fraction of CGB emission that can be resolved into AGN is smaller
than 50\% \citep{2005MNRAS.357.1281W}. This unresolved CGB flux could
be explained by photon emission of a yet undetected population of
highly absorbed AGN. The spectrum of the CGB is
shown in Fig. \ref{fig:earth_albedo_spec} for the spectral
parameters given in \citet{Ajello:2008xb}.\\

\begin{figure}[h]
  \centering
  \includegraphics*[]{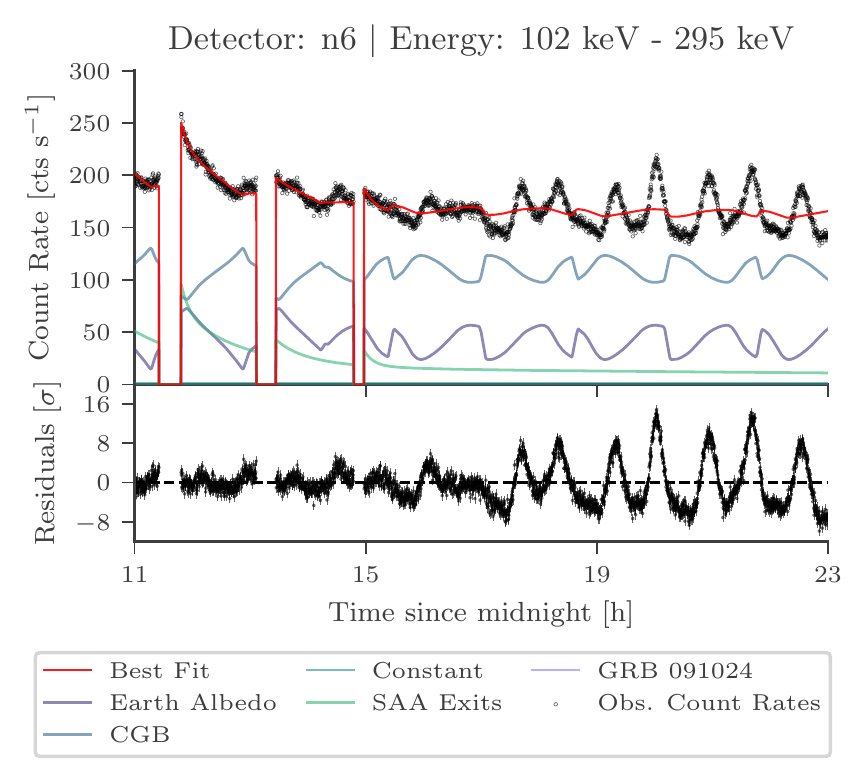}
  \caption{Same as Fig. \ref{fig:counts_earth_albedo} but
    including the Cosmic Gamma-Ray Background as additional background source.
    The influence of the Earth Albedo and the CGB are strongly correlated.
    When the influence of the Earth Albedo is high the Earth is covering a part of the response with
    high effective area and also blocks the CGB photons from this direction and therefore leads to a lower influence of the CGB.}
  \label{fig:counts_cgb}
\end{figure}

Fig. \ref{fig:counts_cgb} shows the model expanded by the
component for the CGB. The assumed functional form of the CGB
spectrum in this paper is a smoothly connected double power law with the spectral
parameter values as given in \citet{Ajello:2008xb}, but we free the normalization
$C$ and fit it to the data (see Eq. \ref{eq:cgb_spec}). \\
It was additionally shown theoretically by \citet{Churazov+2008} that a significant percentage of the CGB spectrum in the keV range gets reflected by the Earth. As this is a second order influence it is currently not included in the work, but we plan to include this reflection of the CGB from the Earth in the future.

\begin{equation}
  \label{eq:cgb_spec}
  \frac{\mathrm{d}N}{\mathrm{d}E} = \frac{C}{ ( \frac{E}{30 \mathrm{keV}}) ^{1.32} + ( \frac{E}{30 \mathrm{keV}})^{2.88}}
\end{equation}

\subsubsection{Cosmic Rays}
\label{cosmic_rays}
Cosmic rays that hit the satellite can either leave a direct signal in
the detectors, that is converted into a certain
PHA channel, or excite atoms in the satellite material that upon
de-excitation produce photons (as shown for INTEGRAL/SPI in
\citet{2018A&A...611A..12D}) which can be measured by GBM's
detectors. Due to Fermi's low-Earth orbit and the small inclination,
the detectors are well protected by the Earth magnetic field against
cosmic rays. Only the cosmic rays with very high
energies (> GeV for protons) can reach the satellite \citep{2009arXiv0912.3611P}.\\
This shielding of cosmic rays varies during the orbital motion with
the different magnetic field strength of the Earth in different orbital
positions and therefore the amount and the spectrum of cosmic rays
hitting the satellite is time dependent.  We will introduce two
different approaches to model the cosmic ray contribution, first by
using the McIlwain L-parameter
and second by the use of the data in the high-energy BGO channels.\\

The McIlwain L-parameter \citep{10.1007/978-94-010-3553-8_4} can be
used as quantifier to model this shielding effect against cosmic rays
by the Earth's magnetic field as has been shown for example for
protons and electrons measured
by the LAT in \citet{2009arXiv0912.3611P}. The McIlwain L-parameter is
    connected to the magnetic cutoff rigidity $R_C$ as given in Eq. \ref{McL_cutoff}.
    The parameters $K$ and $\alpha$
  change with time and have to be determined for the corresponding time \citep{McIlwain}.
\begin{equation}
  \label{McL_cutoff}
R_C = K L^{-\alpha}
\end{equation}
If one extends the magnetic field line at a position to the magnetic
equator, then the L-parameter at that position gives the distance of
the magnetic field line at the magnetic equator to the Earth center in
units of Earth radii.  The correct functional dependence of the
background rate caused by cosmic rays on the L-parameter is not
known. All that is known a priori is that a higher L-parameter means
a weaker protection by the Earth magnetic field and thus should lead
to a higher background caused by cosmic rays. During our work we found
that assuming a linear relationship between the background
rate difference and the L-parameter difference seems to describe the
detected background rates quite well for the GBM detectors and the
L-parameter range on the Fermi orbit ($\approx$ 1-1.7 outside of the
SAA as shown in \citet{2009arXiv0912.3611P}).
\begin{equation}\label{eq:cosmic_ray_diff}
  R(L_1)-R(L_2) \propto L_1 - L_2
\end{equation}

\noindent For cosmic rays it is assumed that Eq.
\ref{eq:cosmic_ray_diff} describes the difference in the count rates
for different L-parameter values. To get the McIlwain L-parameter
values for the different times the weekly spacecraft
file\footnote{from
  heasarc.gsfc.nasa.gov/FTP/fermi/data/lat/weekly/spacecraft/} is
used. Parameter values are provided with a time resolution of
30s. Between these time steps a linear interpolation is used. Due to
the fact that we only have a formula for the difference in the rate
for different L-parameters, the total rate due to the cosmic rays is
defined as
\begin{equation}
  R_{\mathrm{CR}} = R_{\mathrm{CR}}(L_{\mathrm{min}}) + C_{\mathrm{McIlwain}} \cdot (L-L_{\mathrm{min}}).
\end{equation}

\noindent Here $L_{min}$ is the minimal L-parameter of the data. Therefore, the
background model component has two parameters: Firstly, a constant
$R_{CR}(L_{min})$, which is the same for all time bins, and secondly,
a normalization $C_{McIlwain})$, which is multiplied by the difference
of the L-parameter of the time bin and the minimal L-Parameter in the
data. This constant contribution motivates the use of our constant
source we introduced in Sect. \ref{parag:model_const}. The influence
of the cosmic rays in our model is direction independent. This seems
to be true to first order, but we see in the high-energy channels,
where the cosmic ray contribution is dominant, directional differences
of a few tens of percent that can not be explained by our model. This
could be explained by the East-West asymmetry of the cosmic ray flux
due to the Earth magnetic field (see Sec. \ref{earth_albedo}).\\

Because the influence of cosmic rays on GBMs background is rising with
energy, one can assume that in the high-energy channels of the BGO
detectors the background is strongly dominated by cosmic rays. This
can be used as an alternative to model the cosmic ray contribution. To
obtain the functional form, we binned the BGO data in the energy range
of 8.6 MeV to 16.6 MeV to 100s and fit a spline of 3rd degree with a
smoothing prior. It is then assumed that the variation in the amount
of background rates caused by cosmic rays in the lower energies is
linearly correlated with the functional obtained from the BGO data;
therefore, we only fit for the normalization $C_\mathrm{{BGO-Approx}}$
that is multiplied with the fitted spline $S_{\mathrm{BGO}}$.

\begin{equation}
  R_{CR} = C_{BGO-Approx} \cdot S_{BGO}
\end{equation}

\begin{figure}[h]
  \centering
  \includegraphics*[]{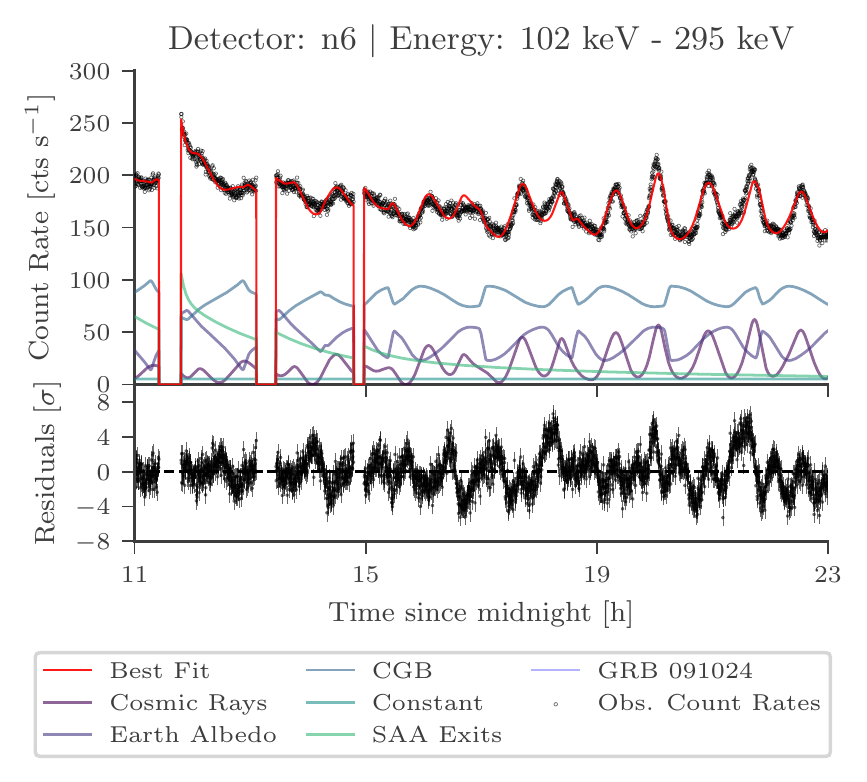}
  \caption{Same as Fig. \ref{fig:counts_cgb}, but including the cosmic
    ray component. For this plot the McIlwain L-parameter model is
    used.  It is clearly visible that the fast raising and lowering of
    the count rates is mainly caused by the difference of cosmic rays
    reaching the detector.}
  \label{fig:cosmic_rays_l}
\end{figure}

\begin{figure}[h]
  \centering
  \includegraphics*[]{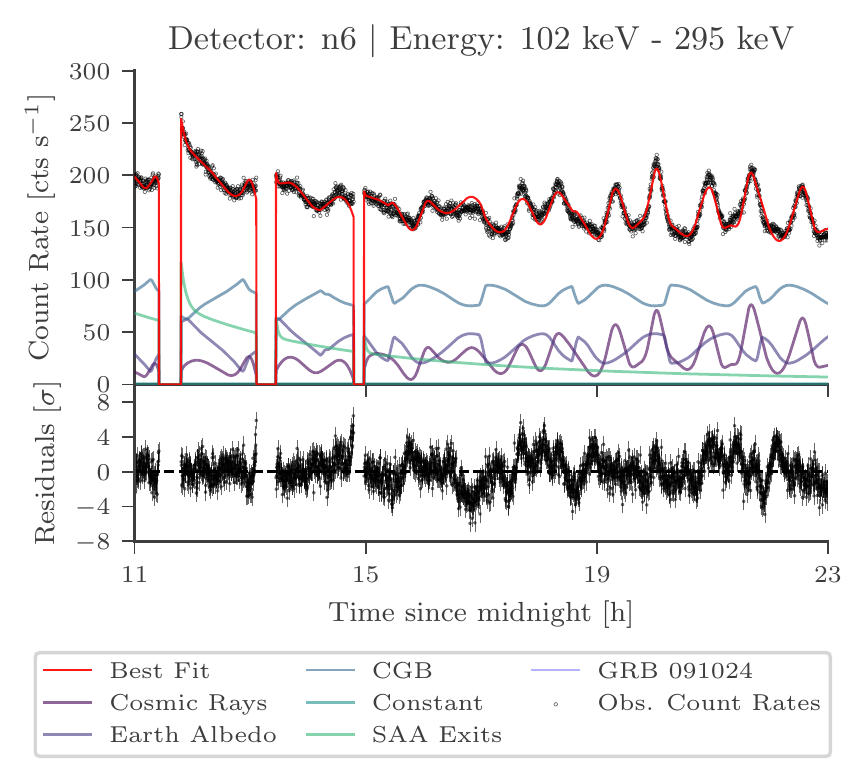}
  \caption{Same as Fig. \ref{fig:counts_cgb}, but including the cosmic
    ray component. For this plot the BGO approximation model is
    used. It is clearly visible that the fast raising and lowering of
    the count rates is mainly caused by the difference of cosmic rays
    reaching the detector. The BGO approximation can fit the data
    noticeably better than the McIlwain L-parameter model (see
    Fig. \ref{fig:cosmic_rays_l}).}
  \label{fig:cosmic_rays_bgo}
\end{figure}

\noindent The model obtained by the BGO approximation
(Fig. \ref{fig:cosmic_rays_bgo}) can fit the cosmic ray contribution
noticeably better than the model that uses the McIlwain L-parameter
(Fig. \ref{fig:cosmic_rays_l}). But of course one has to keep in mind that the BGO
data is not free of other background sources as for example the spectrum of the
Earth Albedo, that was introduced in Sec. \ref{earth_albedo}, extends to the MeV
range. This shortfall could be overcome by the use of the
Anti-Coincidence Detector (ACD) of the LAT to model the cosmic ray background in
GBM. The ACD is primarily used to detected charged particles in 89
tiles around the LAT, in order to exclude the charged particle induced
background in the LAT data. Because the count rates of the ACD are
also stored, one could use the different tiles to reconstruct the
particle flux variation for different sites of the satellite
individually. Unfortunately, the data of ACD is not publicly available,
which is why we leave this approach for future improvements of the
model and use the BGO approximation in the following sections.\\
The BGO approximation is the only empirical model component in this paper.
  This is due to the complex processes that lead to the background contribution by the cosmic rays,
  like their deflection in the earth magnetic field and how they activate the satellite material.
  An appropriate procedure to deal with this would be a full simulation, but unfortunately there is no public mass model of the Fermi satellite that could
  be used to setup such a simulation.

\subsubsection{Point Sources}

\begin{figure}[h]
  \centering
  \includegraphics*[]{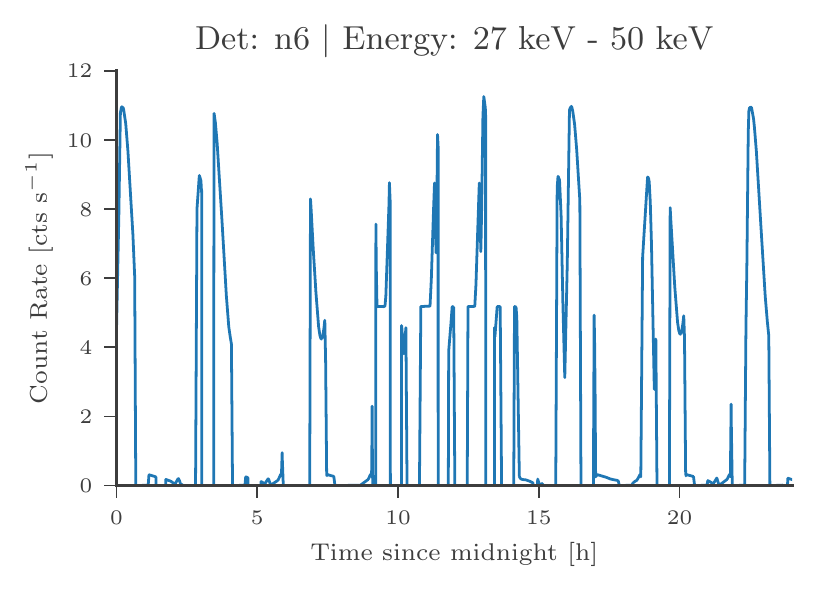}
  \caption{The expected temporal count rate variation for the reconstructed energy
    range 27 to 50 keV caused by the Crab Nebula with a spectrum as given
    in Eq. \ref{eq:crab_spec} with C=9.7 \citep{Madsen:2017vxr}.}
  \label{fig:counts_crab}
\end{figure}

The CGB is not completely homogeneous due to a
plethora of resolved point like sources.
Additionally there are very bright
galactic point sources like the Crab nebula. Their influence is more significant
in the lower energy channels up to 100 keV as most sources are thermal, with a
spectrum rapidly falling with increasing energy.
 These point sources can be seen in the
background of the detectors of GBM as shown in
Fig. \ref{fig:counts_crab}. To reduce the computational time
we only take one point source into account, namely
the Crab, but we plan to add more point sources in the future.
The spectrum of the Crab as
described in \citet{Madsen:2017vxr} is used (see Eq. \ref{eq:crab_spec})
and we again fit the normalization $C$ to the data.

\begin{equation}
  \label{eq:crab_spec}
  \frac{\mathrm{d}N}{\mathrm{d}E} = \frac{C}{ E ^{2.1}}
\end{equation}

\subsubsection{Sun}


The Sun can be treated as a special point-like source, which also
emits X-rays in the low keV range. The spectrum was investigated in
\citet{2007ApJ...659L..77H}, where 2$\sigma$ upper limits are given for the
photon flux of the quiet Sun from 3 keV to 200 keV. Even if these upper
limits would be the actual spectrum of the Sun, the influence of the
Sun would be very small
and, if at all, only visible in the lowest detector channels of GBM.\\

An exception are the eruptions of the Sun, when it violently releases
a large amount of particles and photons with higher energies. While
the particles of these so-called solar flares are deflected by the
Geomagnetic field, the photons reach GBM and are clearly visible in
the data. Information about the flares can be obtained from the yearly
flare files which are produced by the National Center for
Environmental Information (NOAA). The solar flares can be modeled by
an exponential decay function, which is set to zero
up to the time when the flare arrives. To keep the number of parameters
manageable we do not include solar
flares in the model at the moment. If there is a strong solar flare we
simply exclude
this time interval from the analysis. The quiet Sun is modeled as a
simple power law spectrum in the
following.

\subsection{Fitting}
The resulting background model has a total of 50 free parameters for one day, one reconstructed
energy channel and one detector, including the following:
\begin{itemize}
\item Two normalization and two decay constants for every SAA exit
  ($\approx$ 40 parameters in total; depends on the number of SAA exits during the day).
\item Two normalization and two decay constants for leftover
  excitation at the beginning of the day (4 parameters).
\item Normalization for Earth Albedo, Cosmic Gamma-ray Background,
  Sun, point sources (only Crab at the moment), cosmic ray
  contribution (5 parameter).
\item One constant
\end{itemize}
We fitted the free parameters of our model to the data using the
Bayesian nested sampler MultiNest \citep{2009MNRAS.398.1601F,
  2014A&A...564A.125B}.\\

As our model should
represent the total count rates of GBM, we do not have a 'background'
signal. Thus, the correct likelihood is a Poisson likelihood without
background, namely the one given in \citet{1979ApJ...228..939C}:

\begin{equation}
  log(\mathcal{L}) = \sum_{\mathrm{time} \, \mathrm{bins}} \bigg[ N_{\mathrm{model}}(t) - N_{\mathrm{data}}(t) \cdot log(N_{\mathrm{model}}(t))\bigg]
  \label{likelihood}
\end{equation}

The parameter estimation of our model with the data of one day and one
reconstructed energy channel takes $\approx$1:30h with 32 cores (Intel(R) Xeon(R) Platinum 8168 CPU @ 2.70GHz)
producing full posterior distributions. A Bayesian way of
checking the model's ability to describe the observed data are
posterior predictive checks (PPC) \citep{2017arXiv170901449G, 2019MNRAS.490..927B}.
By the use of the posterior predictive distribution

\begin{equation}
  \pi(y^{\mathrm{sim}}|y^{\mathrm{obs}}) = \int \pi(y^{\mathrm{sim}}|\theta) \pi(\theta|y^{\mathrm{obs}}) \mathrm{d}\theta
\end{equation}
\noindent
where $\pi(\theta|y^{\mathrm{obs}})$ is the posterior distribution of $\theta$ and
$\pi(y^{\mathrm{sim}}|\theta)$ the probability of a new simulated observation given a
set of parameters $\theta$, one can generate simulated data and compare it
to the original observed data. We sampled 300 parameter sets from the
posterior distribution and simulated new data for each of these sets
using the underlying Poisson distribution of photon counts. The
observed data should be similar to the simulated data which can
visually be checked by making the area which contains 95\% of the
simulated data and verifying if the observed data lies within. It is a
strong indication that the fit is wrong if many observed data points
lie outside of this marked area. This area that contains 95\% of the
simulated data is marked as green area around the best fit in the
following.\\

The shown Poisson residuals are computed under the hypothesis that
there is no uncertainty in the background. In other words, the
probability of obtaining the observed counts is computed, given the
expected counts from the background and then transforming it in units
of sigma.

\section{Results}
\label{sec:results}
To demonstrate the usefulness of our newly developed background model
in a practical setting we now provide several fits of the model to different days. In \ref{110920A} we show how the model can be used to fit the background for the GRB 110920A for which the classical approach of using polynomial fits can give ambiguous answers. After this we show in \ref{UL_GRB} that the multiple emission peaks over more than 1000 seconds caused by the ultra long GRB 091024 does not effect our fit even if we do not exclude the emission times from the fit. To check this for an even more extreme case we show in \ref{cygni} the background fit for the 21st June 2015, which was during the time of a V404 Cygni outburst. This outburst caused several very long emission periods seen in GBM during this day. We will show that even such an extreme event only slightly disturbs our background fit and that we can identify more emission times than the ones that triggered GBM. In the final example we show that we can identify untriggered excess emission for GRB 130925A with this background model.

\subsection{GRB 110920A}
\label{110920A}
\begin{figure}[h]
  \centering
  \includegraphics*[]{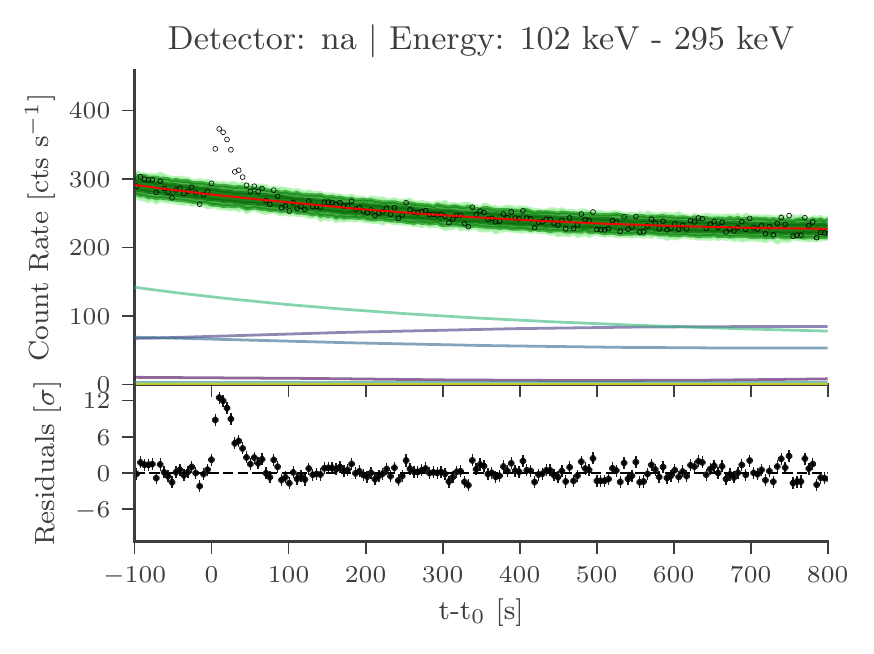}
  \caption{Background fit for the time around GRB 110920A in the reconstructed energy range of 102 to 295 keV for detector na with a time-bin size of 5 seconds. The model can fit the background very well and the GRB is well visible in the residuals.}
  \label{fig:110920A_fit}
\end{figure}

\begin{figure}[h]
  \centering
  \includegraphics*[]{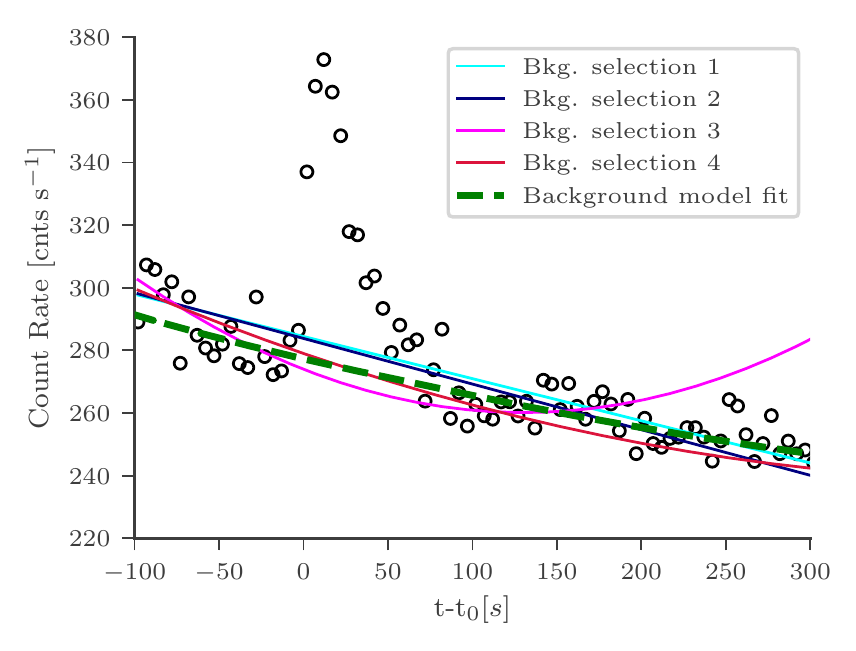}
  \caption{Data and different background estimations for the time around GRB 110920A in the reconstructed energy range from 102 to 295 keV for detector na with a time-bin size of 5 seconds. The four solid lines are derived with the classical approach of fitting polynomials to the time before after the transient event, each with a different definition of these times (1: -100s to -90s and 150s to 200s; 2: -100s to -70s and 50s to 200s; 3: -100s to 0s and 100s to 200s; 4: -100s to -50s and 400s to 500s). Additionally the best fit of our background model is plotted as dashed green line. It is clear that the 4 different polynomial background estimation give different results, that all look equally valid in a short time around the GRB event ($\pm$ 50 seconds), which makes a random choice of one of them necessary, whereas the fitted background model gives an unambiguous answer.}
  \label{fig:110920A_poly}
\end{figure}

GRB 110920A was a bright, single pulse GRB that occurred only about 100 seconds after an SAA exit of GBM. This causes the background to have a significant component from the exponential decay of the activated material. In Fig. \ref{fig:110920A_fit} we show that our model can explain the background around GRB 110920A very well. For an accurate fit of the background during the source time of the GRB event we had to exclude the time of the GRB event from the fit, because it happened so shortly after an SAA exit, that it disturbed the SAA exit fit badly. But we see the GRB as clear deviation from the background fits even without excluding the time, which means that this GRB could have been found with the background model without prior knowledge of the time of the GRB event. \\
In Fig. \ref{fig:110920A_poly} we show the problems that can occur when one is using the classical polynomial approach for a background that can not really be fitted by a polynomial, as it contains a significant component from the exponential decay of the activated material after the SAA exit. The figure shows different polynomial fits, for different background time selections around the GRB event. One can see, that the background fits are quiet different during the active time of the GRB but all look equally valid in a short time around the GRB event ($\pm$ 50 seconds). This random choose of one of these background polynomials can be overcome with our background model, that gives only one unambiguous answer.

\subsection{Ultra-Long GRB 091024}
\label{UL_GRB}
In this subsection we present the background fits for the
24th October 2009, that contains the previously reported ultra-long
GRB 091024 \citep{2011A&A...528A..15G}.
GRB 091024 was detected by Fermi-GBM \citep{GCN_10070}, Swift-BAT \citep{GCN_10062}
and Konus-Wind \citep{GCN_10083} and consisted of multiple emission peaks
with a total duration of about 1000 seconds. The
last emission peak had a duration of about 230 seconds, which makes it
an ideal test case, to evaluate if this long, multiple emission period could have
effected our background fit. \\

\begin{figure}[h]
  \centering
  \includegraphics*[]{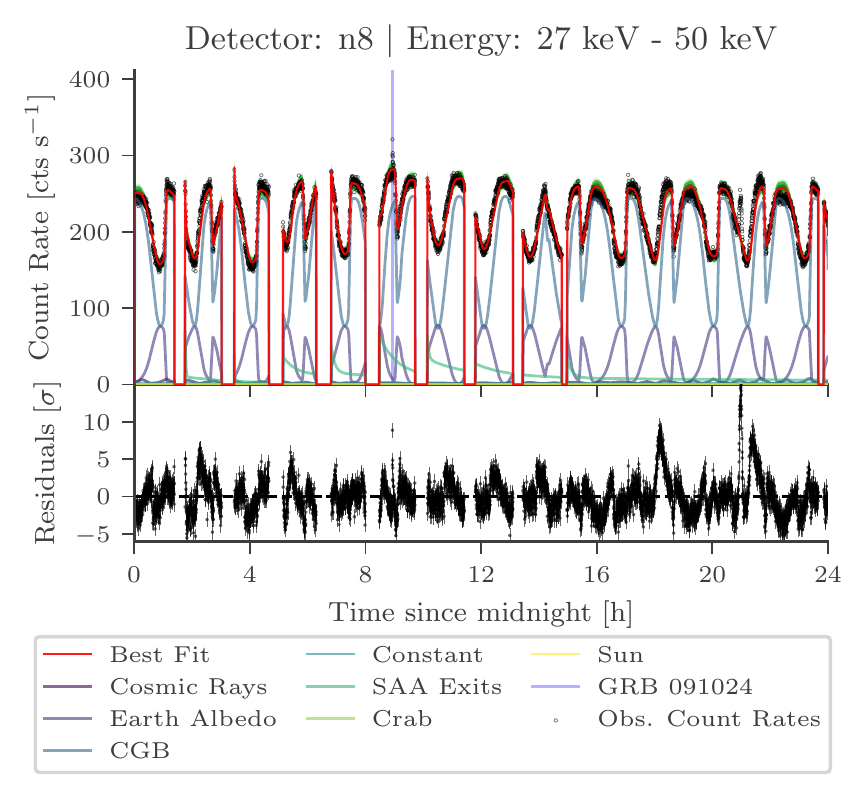}
  \caption{Data and background fit for 24/10/2009 including the
    ultra-long GRB 091024 for detector n8 in the reconstructed energy range 
    of 27--50 keV with a time-bin size of 15\,s. In the low energy range
    the background is dominated by the CGB, but also has some
    significant contribution from the Earth Albedo and the SAA
    exits. The plot shows the influence of the different sources
    in different colors, the best fit total model in red and the
    detected count rates in black. The GBM trigger time of GRB 091024 is marked by a cyan
  line. }
  \label{fig:091026_n8_e2}
\end{figure}

\begin{figure}[h]
  \centering
  \includegraphics*[]{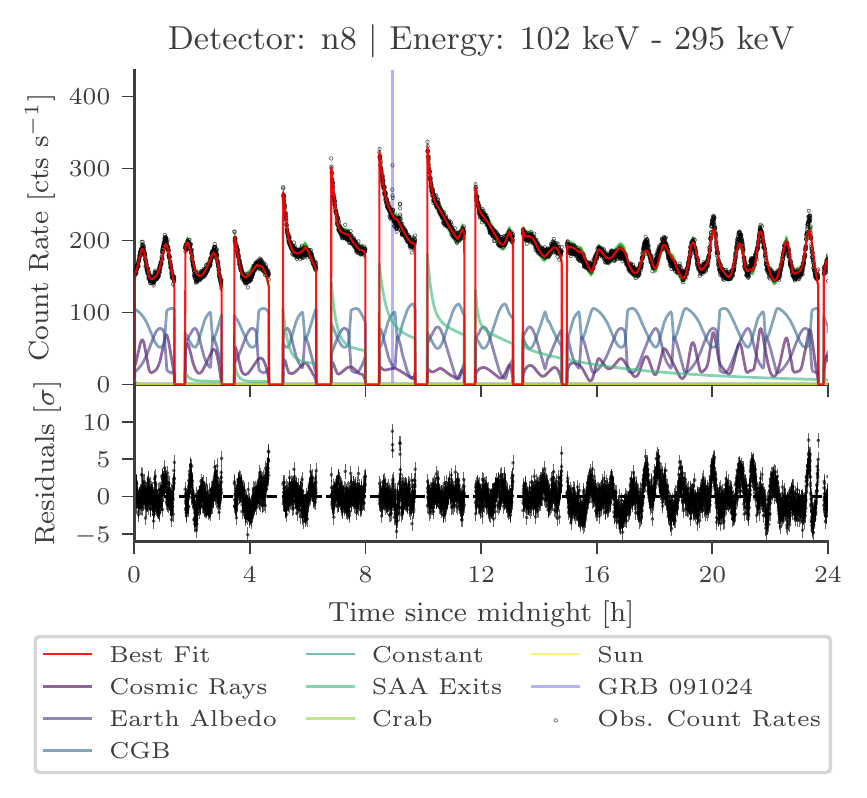}
  \caption{Data and background fit for 24/10/2009 including the
    ultra-long GRB 091024 for detector n8 in the reconstructed energy range of 102--295 keV with a time-bin size of 15\,s. In this reconstructed energy range
    the contribution from the cosmic rays is also important.
    The GBM trigger time of GRB 091024 is marked by a cyan line.}
  \label{fig:091026_n8_e4}
\end{figure}

Fig. \ref{fig:091026_n8_e2} and \ref{fig:091026_n8_e4} show the data
(black) and the best fit of the total background model (red) for the
two reconstructed energy ranges 27--50 keV and 102--295 keV for detector n8.
The lower energy range is, as previously
discussed, dominated by the Cosmic Gamma-ray Background and the Earth
Albedo, while the cosmic ray component is negligible. In the reconstructed 
energy range from 102--295 keV the cosmic ray component is already important.

\begin{figure*}
  \centering

  \begin{subfigure}[b]{0.45\textwidth}
    \centering
    \includegraphics[width=\textwidth]{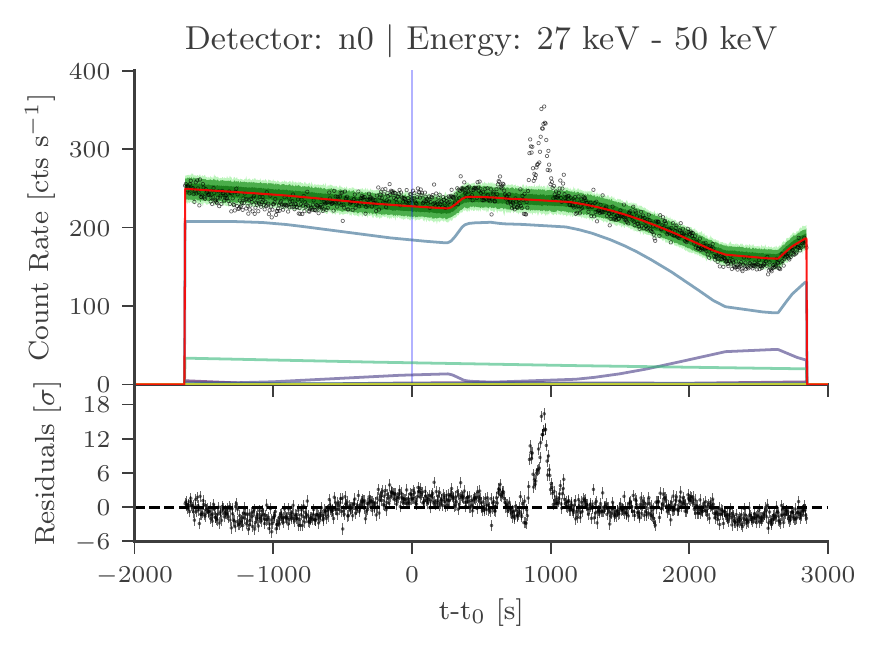}
    \label{fig:shielding_a}
  \end{subfigure}
  \hfill
  \begin{subfigure}[b]{0.45\textwidth}
    \centering
    \includegraphics[width=\textwidth]{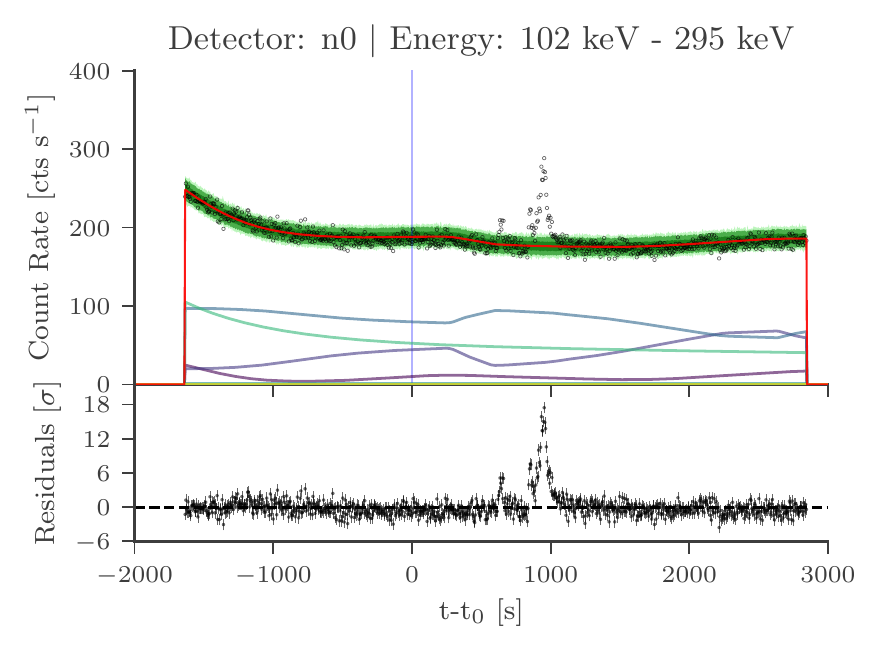}
    \label{fig:shielding_b}
  \end{subfigure}
  \begin{subfigure}[b]{0.45\textwidth}
    \centering
    \includegraphics[width=\textwidth]{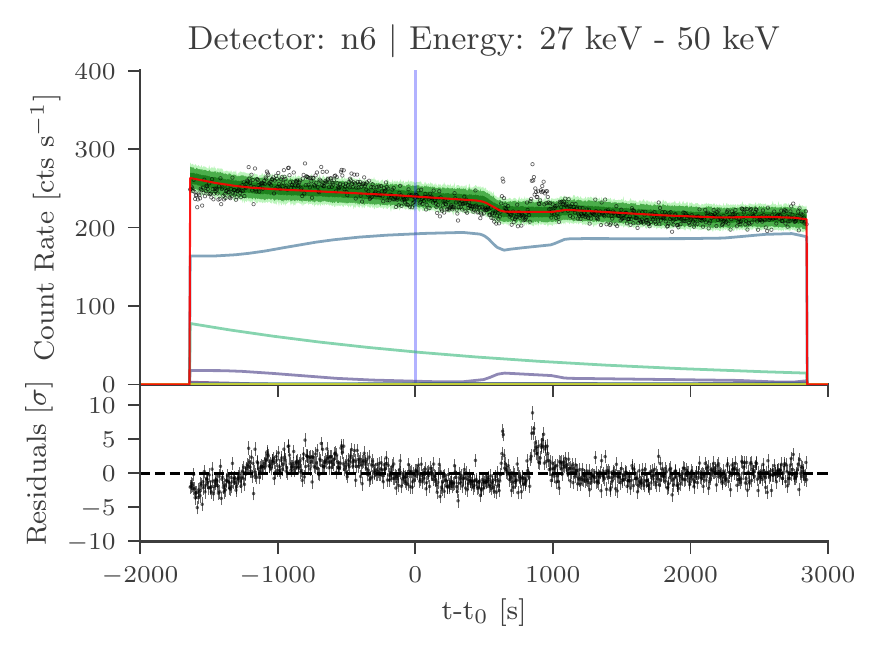}
    \label{fig:shielding_c}
  \end{subfigure}
  \hfill
  \begin{subfigure}[b]{0.45\textwidth}
    \centering
    \includegraphics[width=\textwidth]{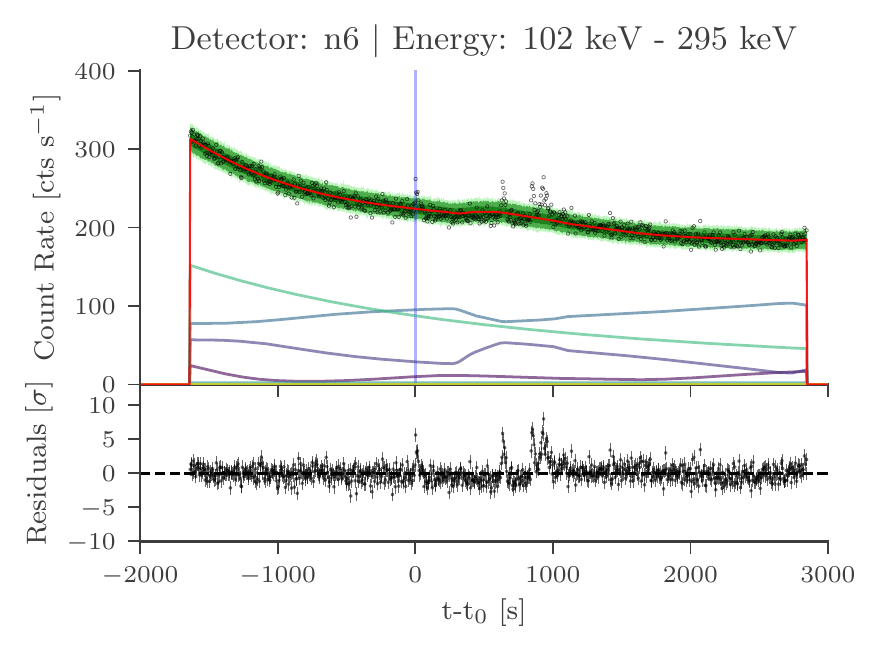}
    \label{fig:shielding_d}
  \end{subfigure}
  \hfill
  \begin{subfigure}[b]{0.45\textwidth}
    \centering
    \includegraphics[width=\textwidth]{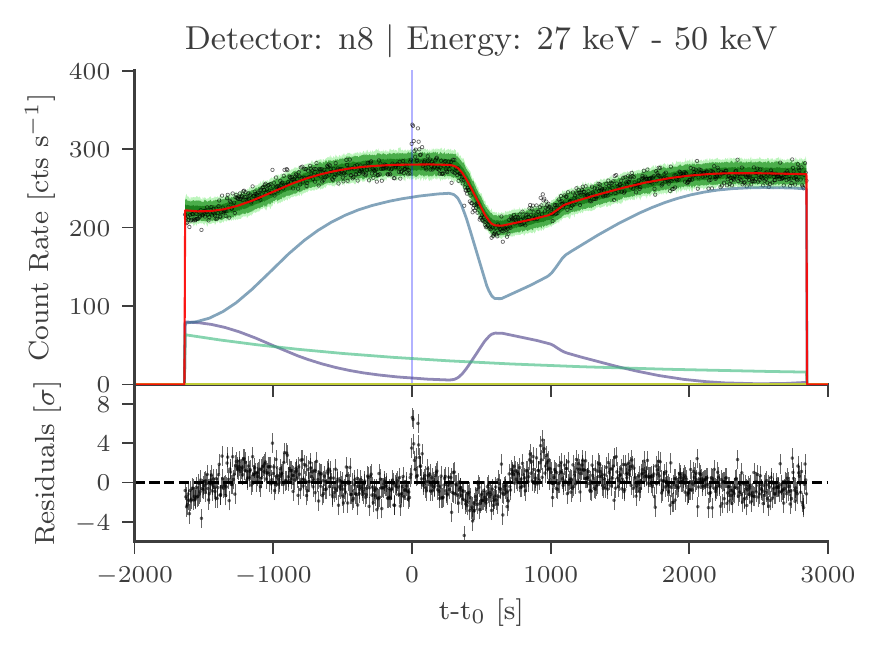}
    \label{fig:shielding_d}
  \end{subfigure}
  \hfill
  \begin{subfigure}[b]{0.45\textwidth}
    \centering
    \includegraphics[width=\textwidth]{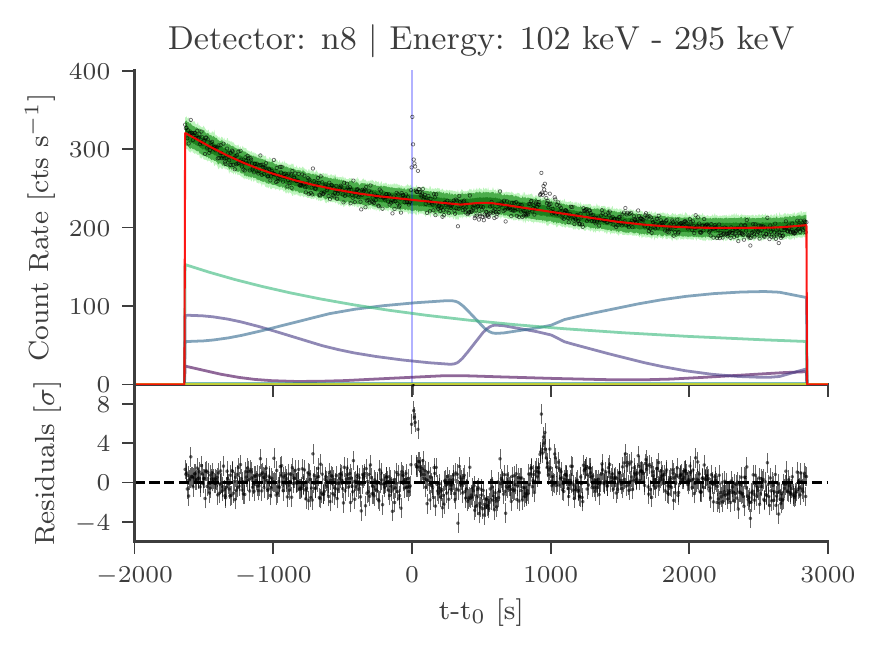}
    \label{fig:shielding_d}
  \end{subfigure}
  \begin{subfigure}[b]{1\textwidth}
    \centering
    \includegraphics[width=\textwidth]{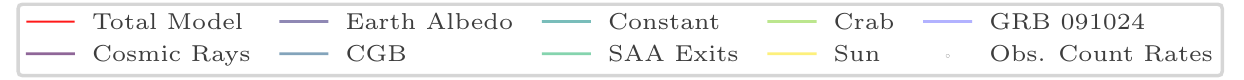}
   \end{subfigure}

   \caption{Data and background fits for the 5000 seconds around the
     GBM trigger time of GRB 091024 for three different detectors
     (n0 (top), n6 (middle), n8 (bottom)) and two reconstructed energy ranges (27-50 keV (left) and
     102-295 keV (right)). The multiple peak emission with a total duration of
     more than 1000 seconds after the GBM trigger (t=0) are clearly visible as deviations from the
     background fit. Due to the long duration of GRB 091024 the
     detector with the smallest offset angle to the GRB position has
     changed from the first to the last peak from n8 to n0, which
     explains why the last peak is clearly visible in the light curves
     of n0 while the first peak is not.}
  \label{fig:grb091024}
\end{figure*}

It is clearly visible in the residuals and with the posterior
predictive check, that the model can  explain most of the observed data 
very well. In both reconstructed
energy ranges an excess of count rate for the trigger time is already
visible in Fig. \ref{fig:091026_n8_e2} and \ref{fig:091026_n8_e4},
but is much more obvious in Fig.  \ref{fig:grb091024}, which shows
the background fit for three different detectors and two different
reconstructed energy ranges around the trigger time of GRB 091024. Here,
the multiple emission peaks are clearly visible as
deviations from the background fit, demonstrating the usefulness of a
physically motivated background model to identify
long-duration emission, and the
possibility to use the fitted background model as background
estimation during the active time of the transient source. \\

We can also see a second strong deviation from the fitted background
model in the low energy range at about 21 hours after
midnight, for which we checked that it is not from the same
location as GRB 091024. This could be due to a soft transient source.\\
Additionally, there are two long-duration, significant (up to 10 sigma) deviations from the fitted background model in the reconstructed energy range 27--50 keV. This could be due to a missing point source in our model, as we see this deviation in three detectors, that point in a similar direction (n6, n7 and n8), at similar (but not equal) times. To remove such  deviations, we plan to include more known point sources in the background model in the future.

\subsection{V404 Cygni flaring}
\label{cygni}
\begin{figure*}
  \centering

  \begin{subfigure}[b]{0.45\textwidth}
    \centering
    \includegraphics[width=\textwidth]{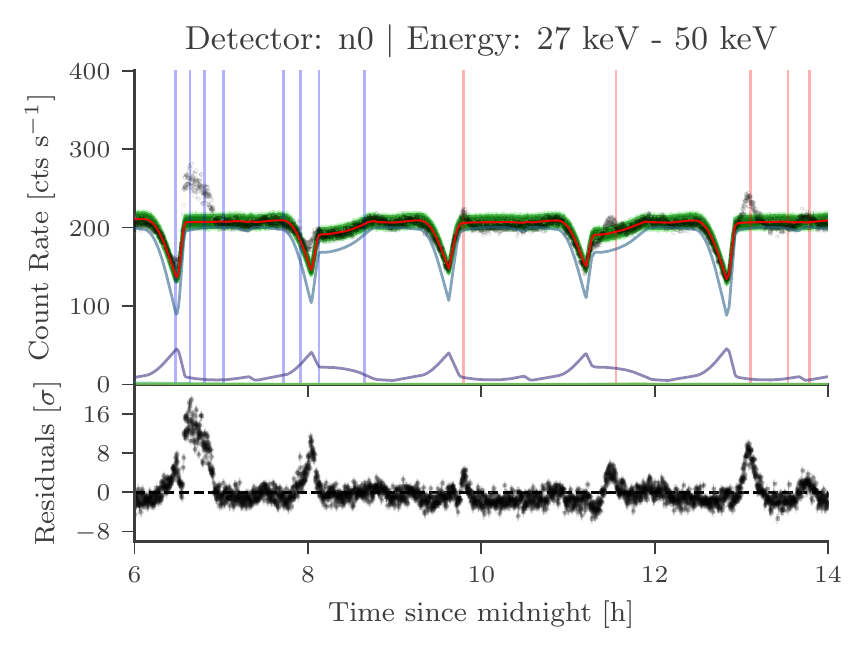}
    \label{fig:shielding_a}
  \end{subfigure}
  \hfill
  \begin{subfigure}[b]{0.45\textwidth}
    \centering
    \includegraphics[width=\textwidth]{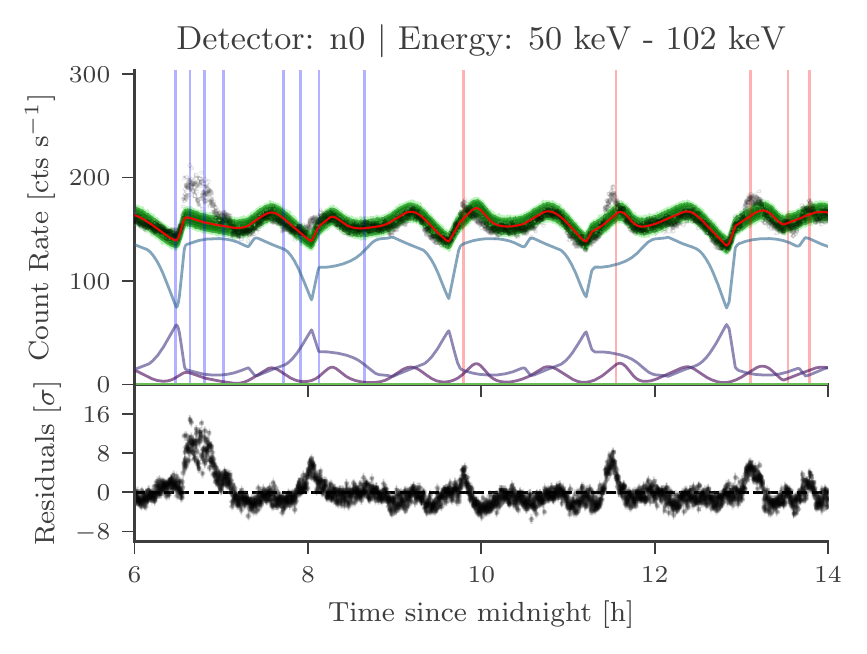}
    \label{fig:shielding_b}
  \end{subfigure}
  \begin{subfigure}[b]{0.45\textwidth}
    \centering
    \includegraphics[width=\textwidth]{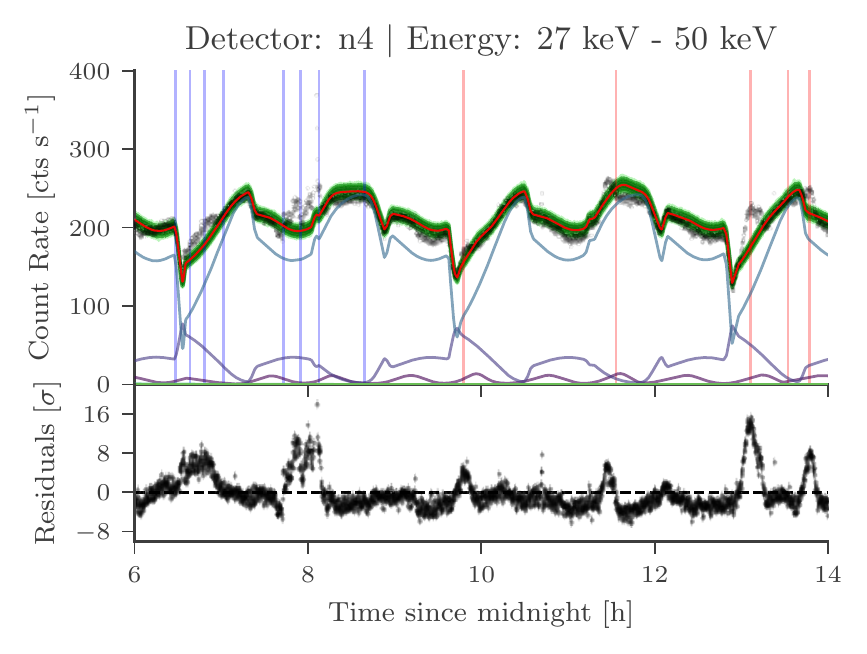}
    \label{fig:shielding_c}
  \end{subfigure}
  \hfill
  \begin{subfigure}[b]{0.45\textwidth}
    \centering
    \includegraphics[width=\textwidth]{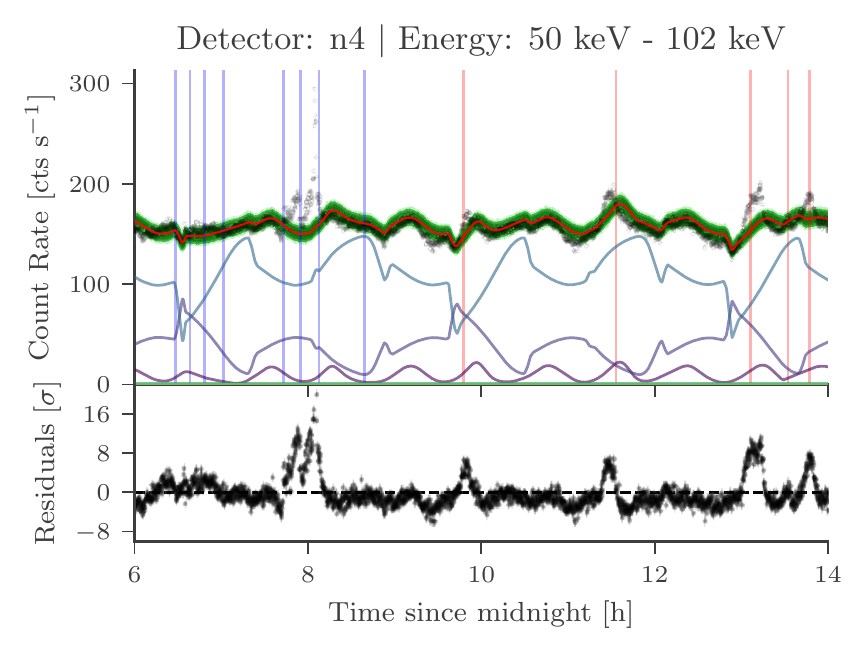}
    \label{fig:shielding_d}
  \end{subfigure}
  \hfill
  \begin{subfigure}[b]{0.45\textwidth}
    \centering
    \includegraphics[width=\textwidth]{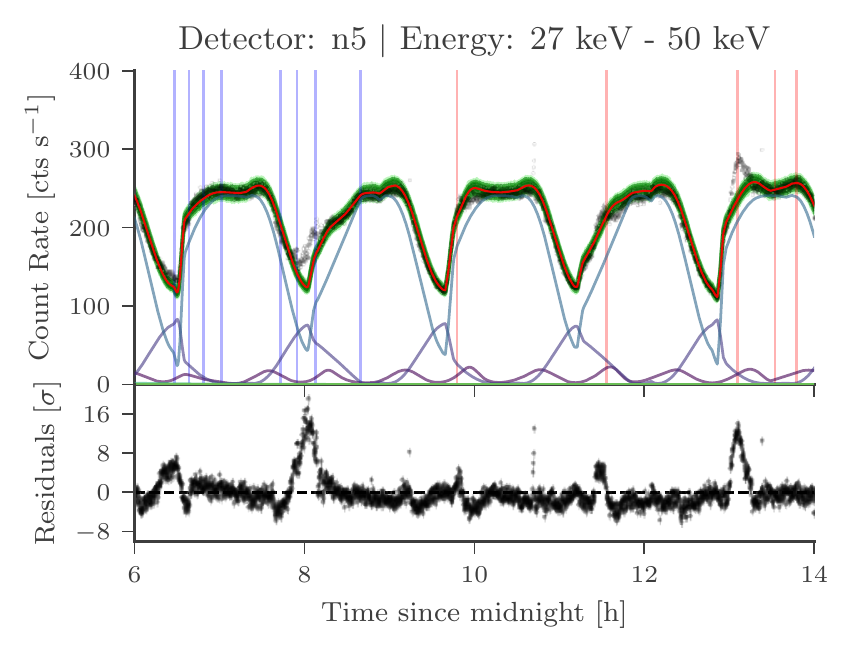}
    \label{fig:shielding_d}
  \end{subfigure}
  \hfill
  \begin{subfigure}[b]{0.45\textwidth}
    \centering
    \includegraphics[width=\textwidth]{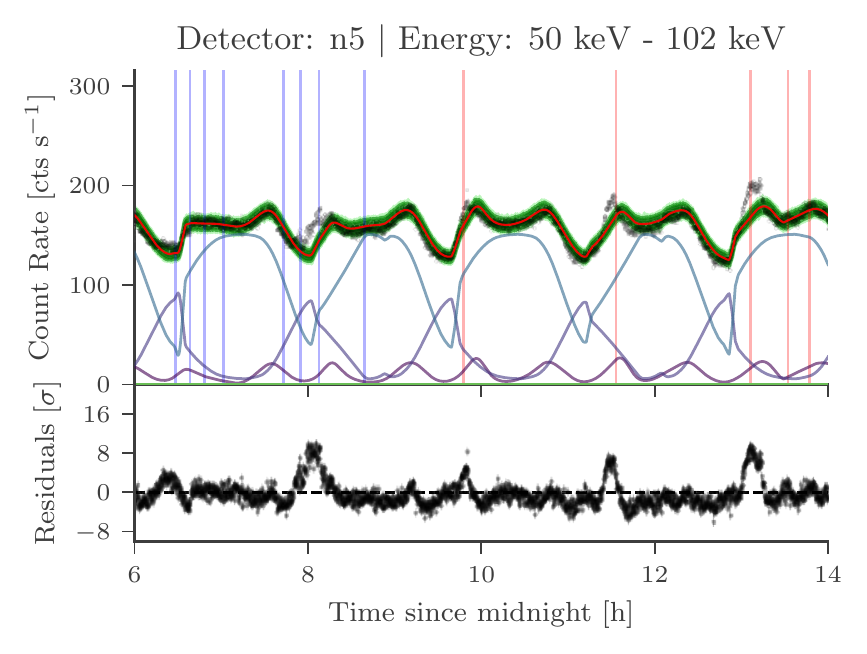}
    \label{fig:shielding_d}
  \end{subfigure}
  \begin{subfigure}[b]{1\textwidth}
    \centering
    \includegraphics[width=\textwidth]{fig/091024/output-min.pdf}
   \end{subfigure}

   \caption{Data and background fits for part of the 21st June 2015 that was during a V404 Cygni outburst
     for three different detectors (n0 (top), n4 (middle), n5 (bottom)) and two reconstructed energy
     ranges (27--50 keV (left) and 50--102 keV (right)). It is clearly visible that there are many
     emission periods during the day that stick out of our background model fit. The blue vertical
     lines are GBM trigger times, whereas the red vertical lines are possible new emission events that
     did not trigger GBM.}
   \label{fig:cygni}

\end{figure*}

In June 2015 V404 Cygni went into outburst, producing multiple flares which
GBM triggered on \citep{GBM_Cygni}.\\
The 21st of June was a very active day, on which GBM triggered 10 times. We present the
light curves of GBM for two energy channels and three detectors for part of the day in Fig. \ref{fig:cygni}. We have marked the times of all GBM triggers with blue vertical lines. One can clearly see from the PPCs and the residuals that there were several more long and weak emission periods on this day, that did not trigger GBM. We marked some of these times roughly with red vertical lines in the plots.
While we have not determined the sky positions of all these
additional emission events, the similarity in the gamma-ray spectrum to that of the triggered events makes the V404 Cyg origin very likely.
V404 Cyg was so active on that day that it disturbed the background fit a bit, as the background fit is slightly too high for the non-active times of V404 Cygni. But still the several emission periods are well visible in the plots even though we did not exclude the times when V404 Cygni was active from the fit. This is due to the fact that we use physical derived background sources and not arbitrary polynomials and therefore very strongly restrict what kind of shapes can be fitted by our background model. This makes the presented background model a promising tool to search for long source emissions in the data of GBM in the future. \\
If one wants to use the fitted background model as an accurate background estimation during the active times of the source, one should of course redo the analysis and exclude the times when the source was active from the fit, to get rid of this disturbance.\\

\subsection{Identifying untriggered excess emission}

\begin{figure}[ht]
  \centering
  \resizebox{\hsize}{!}{\includegraphics{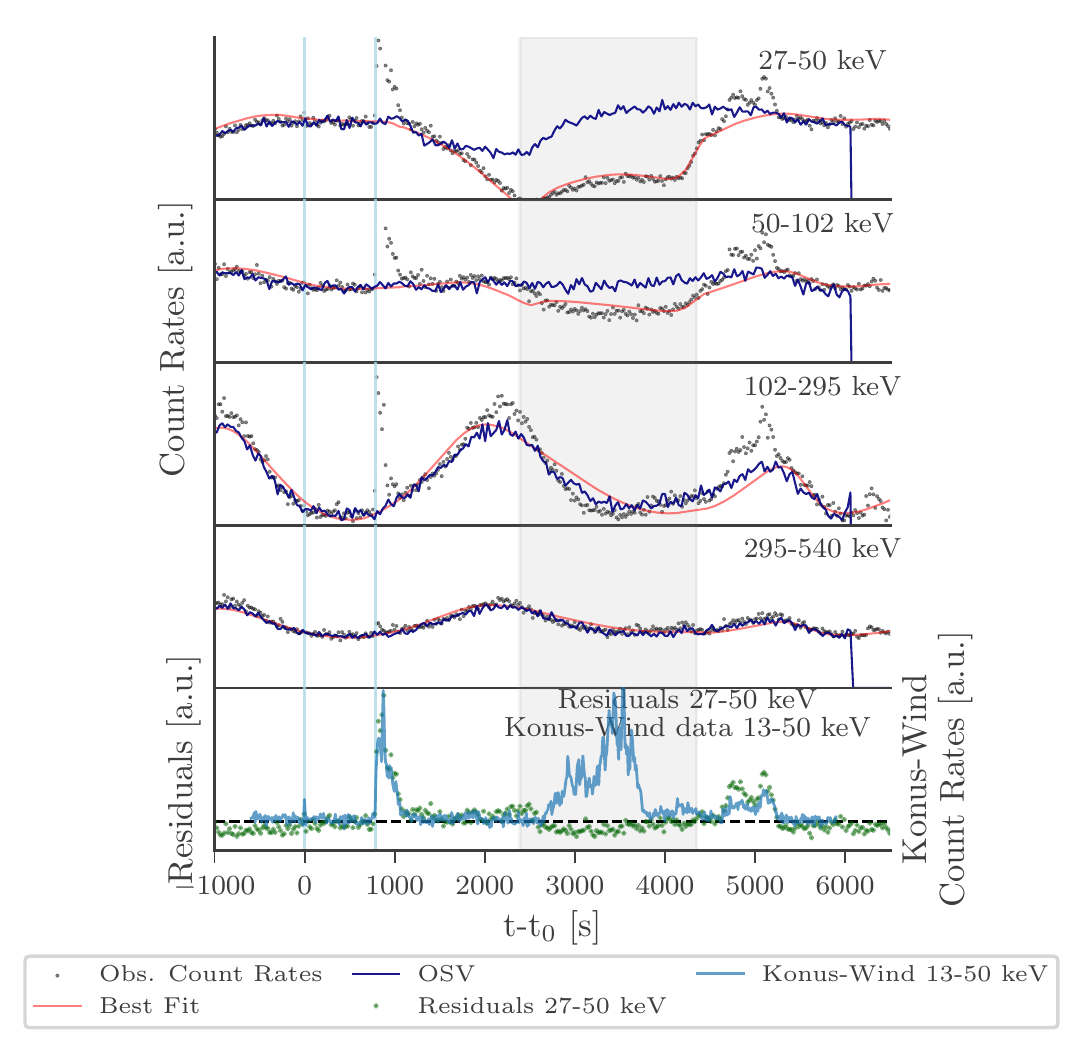}}
  \caption{Data (black), background fit (red) and background estimation (OSV, dark blue) with the \grqq Fermi GBM Orbital Background Subtraction Tool\grqq{} \citep{Fitzpatrickbackground} for the emission time of 130925A, detector n9 and four reconstructed energy ranges in the top four panels. In the lower panel the residuals for the background fit between 27 and 50 keV reconstructed energy are given, as well as the background subtracted Konus-Wind data between 13 and 50 keV \citep{GCN_15260}. The two blue vertical lines show the time of the two GBM triggers for GRB 130925A and the grey area the time during which the position of GRB 130925A was occulted by the earth for GBM. The untriggered excess emission at $\approx$ 5000 seconds after the first GBM trigger is clearly visible in the displayed residuals of the background fit, and fit temporally nicely to the data of Konus-Wind.}
  \label{fig:130925_all}
\end{figure}

Since our background model is fit over a full day, the fit parameters
have no room to allow variations between individual satellite orbits,
i.e. to accommodate spatially (e.g. transient sources) or temporally
(e.g. disturbances in the magnetic field) varying sources. This makes
the background fit very stiff, and thus allows us to readily identify
variable sources.
This is demonstrated for September 25, 2013, 
when multiple emission periods were detected over several thousand seconds  
from the ultra-long GRB 130925A \citep[e.g.][]{GCN_15260}.
In Fig. \ref{fig:130925_all} the data for detector n9 and four reconstructed
energy ranges are displayed, as well as our fitted model (red).
The two GBM triggers are marked as vertical lines.
The other emission period(s), nicely covered by Konus-Wind, verify
the identification of late GRB 130925A emission with GBM.

\subsection{Comparison to \grqq Fermi GBM Orbital Background Subtraction Tool \grqq}

\begin{figure}[h]
  \centering
  \includegraphics*[]{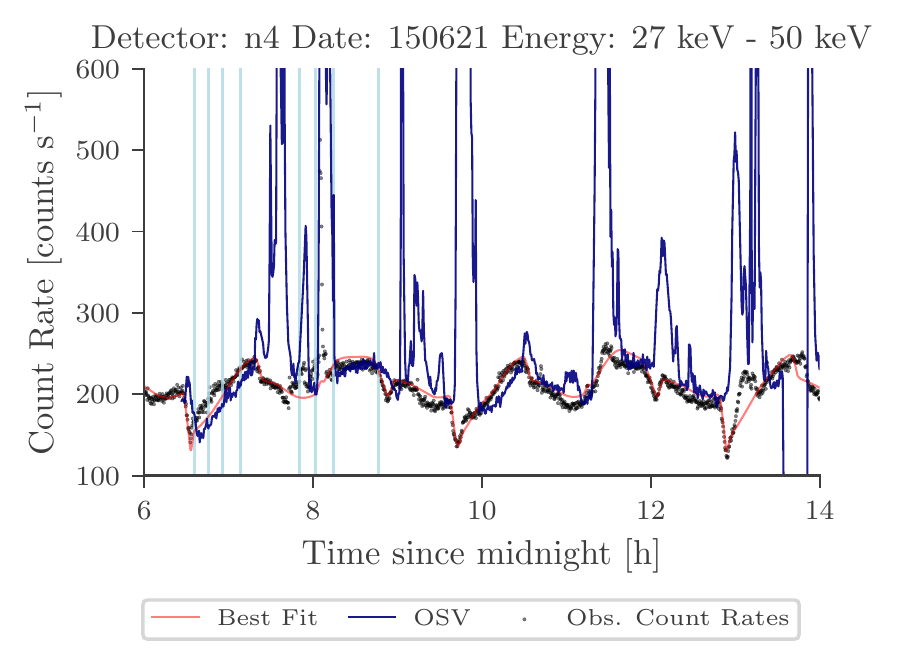}
  \caption{Data (black), background fit (red) and background estimation (OSV, dark blue) with the \grqq Fermi GBM Orbital Background Subtraction Tool\grqq{} for part of 21st June 2015 during the 2015 V404 Cygni outburst, detector n9 and reconstructed energy between 27 and 50 keV. One can see that the OSV estimation fails badly due to the many strong and long emission periods that V404 Cygni produced 30 orbits before and after the shown time, whereas our model fits can explain the background well. All GBM triggers during this time are shown as blue vertical lines (before the onboard triggers were turned off).}
\label{fig:cygni_compare}
\end{figure}

\begin{figure}[h]
  \centering
  \includegraphics*[]{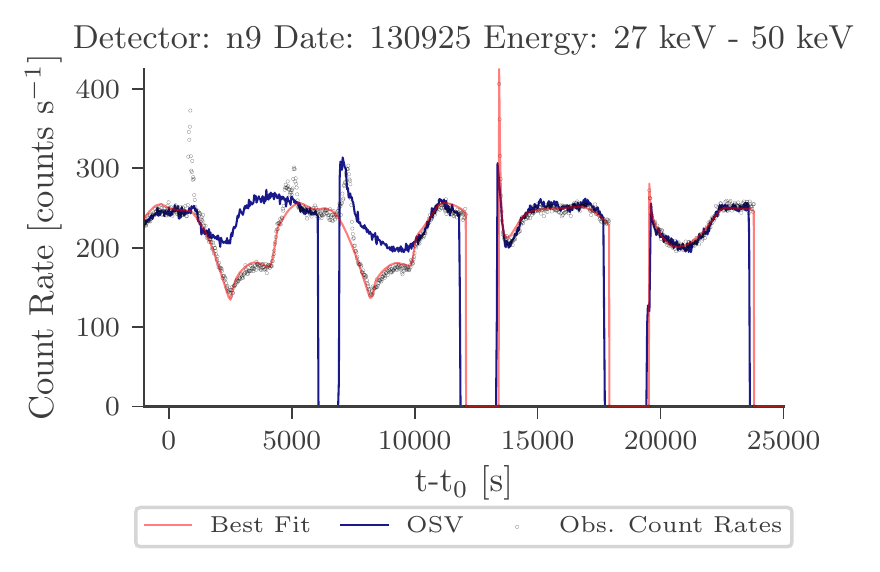}
  \caption{Data (black), background fit (red) and background estimation (OSV, dark blue) with the \grqq Fermi GBM Orbital Background Subtraction Tool\grqq{} for 25000 seconds on the 25th September 2013. One can see that the OSV estimation fails during the first 10000 seconds due to an ARR but gets better after the satellite returned to normal pointing mode. But it does not get better than our model, which can fit all the data well. And for about 1000 seconds between $\approx$ 6000 and $\approx$ 7000 seconds the \grqq Fermi GBM Orbital Background Subtraction Tool\grqq{} can not estimate the background at all, due to a SAA passage 30 orbits before or after this time. The high excess of data above the background fit at $\approx$ 7000 seconds is well visible in the energy channels up to 100 keV in several detectors and will be investigated in the future.}
\label{fig:long}
\end{figure}

The advantage of the presented background model with respect to the 
\grqq Fermi GBM Orbital Background Subtraction 
Tool\grqq\footnote{\url{https://fermi.gsfc.nasa.gov/ssc/data/analysis/user/osv_1.3.tar}} (OSV) 
\citep{Fitzpatrickbackground} is also demonstrated for September 25, 2013
(Fig. \ref{fig:130925_all}), where the estimation of the OSV is overplotted in dark blue.
One can see that our model is capable of explaining the background variations 
in all four reconstructed energy ranges, whereas the OSV fails, mainly due to an Autonomous Repoint Request (ARR) after the second GBM trigger.
But also during days without ARR, the OSV method has its limitations 
(Fig. \ref{fig:cygni_compare} and \ref{fig:long}),
as it depends on having no deviation of any kind (besides ARR also transients
etc) also 30 orbits before and after the day in question. With about
1 onboard trigger every 10 orbits (\url{https://gammaray.nsstc.nasa.gov/gbm/}),
this limitation affects the background estimation for some period of time
on $\sim$30\% of all days.

The following summarizes the advantages of our physical model:
\begin{itemize}
\item is not affected by ARR or other deviations from the normal pointing mode,
\item is not affected if the satellite was in an SAA 30 orbits before or after the time of interest,
\item is not affected by GRBs or any transient sources occurring 30 orbits before or after the time of interest,
\item is a more robust method that does, for example, not need the radiation environment or magnetic field to be stable for at least 3 days (due to the +/- 30 orbits that are used in the OSV),
\item treats the data in a statistically correct way which allows to derive proper errors on the background counts,
\item allows us to understand and study the different components of the background (also interesting for future missions).
\end{itemize}

\section{Conclusion}

The derivation of a physically motivated background model for GBM was presented.
It consists of six different source components (Earth Albedo, Cosmic Gamma-ray background,
  Sun, point sources (just the Crab at the moment), South Atlantic Anomaly and cosmic rays), where only the
cosmic-ray contribution is modeled empirically.\\ 
The presented model is capable of modeling the background for most of
the GBM PHA channels as shown for the example cases of the ultra-long
GRB 091024, the V404 Cygni outburst in June 2015
and GRB 110920A.\\

There are three obvious possible future applications
of this physically motivated background model:
\begin{enumerate}
\item Use it as a new approach
  to search for ultra-long GRBs in the 12+ years of GBM data,
that until now have been indistinguishable from the background
variations.
\item Infer information about the different background components, for example the
Cosmic Gamma-ray Background spectrum in the GBM energy range.
\item Develop a transient search algorithm for slowly rising or long
transients which are not recognized by the onboard trigger algorithm.
\end{enumerate}

\bibliographystyle{aa}
\bibliography{Biltzinger_bkg}

\end{document}